\documentclass[article, prd,nofootinbib,floats,superscriptaddress,eqsecnum,tightenlines,11pt]{revtex4}

\usepackage{hyperref}
\usepackage{graphicx}
\usepackage{amsmath,amssymb,amsfonts,amsthm,latexsym,stmaryrd,bbm}
\usepackage{marginnote}
\usepackage[normalem]{ulem}
\usepackage{color}
\usepackage{braket}
\usepackage{enumerate}
\usepackage{float}

\def\be{\begin{equation}}
\def\ee{\end{equation}}
\def\bea{\begin{eqnarray}}
\def\eea{\end{eqnarray}}
\def\beq{\begin{eqnarray}}
\def\eeq{\end{eqnarray}}
\def\bas{\begin{subequations}\begin{eqnarray}}
\def\eas{\end{eqnarray}\end{subequations}}
\def\nn{\nonumber}

\def\eps{\varepsilon}

\def\tr{\text{tr}}
\def\la{\langle}
\def\ra{\rangle}

\def\f{\frac}

\def\SU{\text{SU}}
\def\SO{\text{SO}}

\def\SL{\text{SL}}
\def\su{\mathfrak{su}}

\def\sl{\mathfrak{sl}}

\newcommand{\C}{{\mathbb C}}
\newcommand{\N}{{\mathbb N}}
\newcommand{\R}{{\mathbb R}}
\newcommand{\Z}{{\mathbb Z}}

\newcommand{\cG}{{\mathcal G}}

\newcommand{\cJ}{{\mathcal J}}
\newcommand{\cK}{{\mathcal K}}

\newcommand{\cH}{{\mathcal H}}

\newcommand{\cD}{{\mathcal D}}
\newcommand{\cC}{{\mathcal C}}

\newcommand{\mat} [2] {\left ( \begin{array}{#1}#2\end{array} \right ) }

\def\tgamma{\tilde{\gamma}}
\def\tc{\tilde{c}}
\def\tp{\tilde{p}}
\def\tv{\tilde{v}}
\def\tb{\tilde{b}}
\def\vj{\vec{\j}}
\def\veta{\vec{\eta}}
\def\bz{\bar{z}}
\def\vtau{\vec{\tau}}
\def\pp{\partial}
\def\rd{\textrm{d}}

\def\ka{\kappa}

\def\eps{\epsilon}

\newcommand{\id}{\mathbb{I}}

\def\balpha{\bar{\alpha}}
\def\bbeta{\bar{\beta}}
\def\vJ{\vec{J}}

\def\vtau{\vec{\tau}}

\def\hJ{\hat{J}}
\def\hK{\hat{K}}
\def\hV{\hat{V}}
\def\hj{\hat{j}}
\def\hk{\hat{k}}

\begin{document}

\title{The Thiemann Complexifier and the CVH algebra \\ for Classical and Quantum FLRW Cosmology}
%\title{The Immirzi Complexifier in Loop Cosmology}
%\title{The Immirzi transform as a unitary operator in Loop Cosmology}

\author{{\bf Jibril Ben Achour}}\email{}
\affiliation{Center for Field Theory and Particle Physics, Fudan Univ., 200433 Shanghai, China}
%\affiliation{Center for Field Theory and Particle Physics, Fudan University, 200433 Shanghai, China}

\author{{\bf  Etera R. Livine}}\email{etera.livine@ens-lyon.fr}
\affiliation{Univ Lyon, Ens de Lyon, Univ Claude Bernard Lyon 1, CNRS, LPENSL, 69007 Lyon, France}
%\affiliation{Univ Lyon, Ens de Lyon, Universit\'e Claude Bernard Lyon 1, CNRS,  LPENSL, 69007 Lyon, France}
\affiliation{Perimeter Institute for Theoretical Physics, Waterloo, Ontario, Canada N2L 2Y5}
%\affiliation{Perimeter Institute for Theoretical Physics, 31 Caroline St North, Waterloo ON, Canada N2L 2Y5}

\date{\today}

\begin{abstract}

In the context of Loop Quantum Gravity (LQG),
 we study the fate of Thiemann complexifier in homogeneous and isotropic FRW cosmology. The complexifier is the dilatation operator acting on the canonical phase space for gravity and generates the canonical transformations shifting the Barbero-Immirzi parameter.
We focus on the closed algebra
%for canonical general relativity
consisting in the complexifier, the 3d volume and the Hamiltonian constraint, which we call the CVH algebra. In standard cosmology, for gravity coupled to a scalar field,  the CVH algebra is identified as a $\su(1,1)$ Lie algebra, with the Hamiltonian as a null generator, the complexifier as a boost and the $\su(1,1)$ Casimir given by the matter density.
%no physical relevance of Immirzi paramter, (Hamiltonian constraint simply gets rescaled)
%In the loop gravity cosmology approach, the curvature is regularized as a $\SU(2)$ holonomy and the cosmological Hamiltonian is accordingly modified.
 The loop gravity cosmology approach introduces a regularization length scale $\lambda$ and regularizes the gravitational Hamiltonian in terms of $\SU(2)$ holonomies. We show that this regularization is compatible with the CVH algebra, if we suitably regularize the complexifier and inverse volume factor. The regularized complexifier generates a generalized version of the Barbero's canonical transformation which reduces to the classical one when $\lambda \rightarrow 0$.
% Then the $\su(1,1)$ Casimir is given by the matter density and { \Ji the minimal volume.}
This structure allows for the exact integration of the actions of the Hamiltonian constraints and the complexifier.
This straightforwardly extends to the quantum level: the cosmological evolution is described in terms of $\SU(1,1)$ coherent states and
%the Immirzi transform, generated by the regularized complexifier, is a unitary operator.
the regularized complexifier generates unitary transformations.
The Barbero-Immirzi parameter is to be distinguished from the regularization scale $\lambda$, it can be rescaled unitarily and the Immirzi ambiguity ultimately disappears from the physical predictions of the theory.
Finally, we show that the complexifier becomes the effective Hamiltonian when deparametrizing the dynamics using the scalar field as a clock, thus underlining the deep relation  between cosmological evolution and scale transformations.

\end{abstract}

\maketitle

\tableofcontents
\newpage

%\bigskip

%%%%%%%%%%%%%%%%%%%%%%%%%%%%%%%%%%

Loop Quantum Gravity (LQG) proposes a background independent framework to quantize general relativity, with a kinematical Hilbert space of quantum states of geometry \cite{LQGBook, LQGReport} and dynamics defined by spinfoam amplitudes. 
Its starting point is the Ashtekar-Barbero  reformulation of General Relativity (GR) as a $\SU(2)$ gauge theory, where the gravitational degrees of freedom are encoded in a canonical pair of conjugate fields, a $\su(2)$-connection $A_{\gamma}$ and the triad $E$.
%$E_{\gamma}$
These are derived by a canonical transformation of the standard canonical pair of the first order formulation of GR given by the extrinsic curvature $K$ and the triad $E$. Indeed the Ashtekar-Barbero connection is defined as $A_{\gamma} = \Gamma[E] + \gamma K$, where $\Gamma[E]$ stands for the  Levi-Civita connection (or rotational spin connection) and $\gamma$ is the Barbero-Immizi (BI) parameter. The generator of this canonical transformation is called the Thiemann complexifier \cite{ThiemannRC} and the BI parameter $\gamma$ is a priori free to take any complex value  \cite{Immirzi1}.
The LQG program then proceeds  to the quantization of these phase space in terms of wave-functions of the Ashtekar-Barbero connection $A_{\gamma}$, with the triad field quantized as a differential operator. 

Trying to find a consistent and clear interpretation of the Barbero-Immizi (BI) parameter in LQG has been challenging. Its status as a new fundamental constant still remains unclear and its puzzling role  in the dynamics of quantum geometry has generated a debate concerning its interpretation which is still open. The BI parameter was initially set to a purely imaginary value, $\gamma=\pm i$, defining the   Ashtekar connection as the (anti-)self-dual Lorentz connection. This nevertheless leads to reality conditions, which we still do not known how to implement at the quantum level. It then appeared that taking real values for the BI parameter, $\gamma\in\R$, allows to develop quantum theory without reality conditions but at the price of foliation-independence. Indeed,  contrary to the self-dual connection,  a real $\su(2)$-connection $A_{\gamma}$ is not anymore a true space-time connection\footnotemark{}  \cite{Samuel}. This is the standard choice for defining loop quantum gravity.
\footnotetext{
The choice of gauge fixing, breaking the local Lorentz invariance under $\SL(2,\C)$ down to its $\SU(2)$ subgroup, is very well understood \cite{BeSa}. There has been extensive work on the definition of $\SL(2,\C)$ connections  and their relation to the Ashtekar-Barbero connections \cite{Alex1LLQG,Alex3LLQG}
%\cite{Alex0LLQG,Alex1LLQG,Alex2LLQG,Alex3LLQG}
as well as investigation on how to embed them back in an explicitly covariant setting \cite{Alex4LLQG,Liv1LLQG,Cianfrani1,MG0,Geiller1LLQG,Geiller2LLQG,JBA1,JBA2}.
%\cite{Alex4LLQG,Liv1LLQG,Cianfrani1,MG0,Geiller1LLQG,Geiller2LLQG,JBA1,JBA2}.
This interplay between $\SU(2)$-invariant quantum states and $\SL(2,\C)$-invariant space-time structures is actually at the heart of the construction of the EPRL spinfoam path integral for the LQG dynamics \cite{Engle:2007wy,Ding:2010ye,Dupuis:2010jn}. Note also that, recently, a new formulation of GR based on time-like hypersurfaces, and using  $\SU(1,1)$ as gauge group, has been developed \cite{KarimLiu}, underlining the independence between the Immirzi ambiguity and the initial choice of gauge fixing. 
}

Pinning down one specific role for the Barbero-Immirzi parameter is impossible. Generated by a canonical transformation, it appears as a coupling constant in the Holst action \cite{Holst}. It can be interpreted as the $\theta$-ambiguity of QCD \cite{YMThetaAmbiguity, ImmirziTop}. It controls the coupling of fermions to the gravitational background  \cite{ImmirziFermion0, ImmirziFermion1, ImmirziPTSym0} and more generally the quantum fluctuations of the torsion field  \cite{ImmTorsion}.
On the one hand, it sets the value of the area and volume minimal eigenvalues: these area and volume gaps define the fundamental scale of the quantum geometry. This is reinforced by its role of a (extrinsic) curvature cut-off which appears to regularize the theory \cite{CharlesCutOff}.
On the other hand, as a coupling constant, we naturally expect it to run with the renormalisation group flow \cite{Benedetti:2011nd}.
It enters explicitly the spinfoam path integral amplitudes (through the simplicity constraints) \cite{Engle:2007wy} while it has been argued to drop out of the symplectic structure after a careful analysis of the LQG  discretization scheme  \cite{Biancha}.
Finally, most predictions from LQG or effective loop gravity models are all $\gamma$-dependent, best illustrated by the maximal density of the bouncing universe in Loop Quantum Cosmology (LQC) \cite{LQCReport, LQCReview, LQCImproved} or the entropy of isolated horizon \cite{EntropyIH2}
%\cite{EntropyIH1, EntropyIH2, LBHStateCounting}
and the resulting black hole thermodynamics \cite{PerezMu4}.

The many faces of the BI parameter suggests two scenarios. Either $\gamma$ plays  the technical role of a regulator in the quantization process and disappear from the physical predictions at the end of the day.
Or $\gamma$ is a fundamental new constant signaling an anomaly, or physically-meaningful deformation, of the quantum algebra of constraints.
In the path to resolving this dilemma, some concrete questions are to be addressed:
\begin{itemize}

\item Are we restricted to real values $\gamma\in\R$ or should we allow for arbitrary complex values $\gamma\in\C$? In particular, should we wick-rotate the theory back to the self-dual case $\gamma=\pm i$ as proposed in \cite{JBA-AC1, JBA-AC2, JBA-AC3}.

\item Is the Barbero canonical transformation to be implemented in a non-unitary way as in the standard LQG formulation \cite{RovThie} or should we look for a unitary representation?

\item Since $\gamma$ fixes the fundamental gap of the  geometric observables, it is hard to envision how it could be renormalized. But  should the dynamics relax this situation (see e.g. the discussion in \cite{Dittrich:2007th,Rovelli:2007ep}) and let the Immirzi parameter $\gamma$ run with the renormalization group and possibly get washed out ?

\end{itemize}

The highly symmetry-reduced context of cosmology provides a perfect arena to analyze in detail the role(s) of $\gamma$. So the goal of this paper is to study the relevance of the BI parameter and test its various meanings in the simplified context of Loop Quantum Cosmology (LQC).
%
%Note that different proposals for constructing the self dual version of LQC have been put forward recently \cite{JBACosmo, EdSDCosmo1, EdSDCosmo2}, but in this paper, we will focus on the real theory, where $\gamma \in \mathbb{R}$.
%
%See \cite{CovSpher, CovGowdy, CovGen, CovBH, CovSG} for recent studies concerning the covariance of symmetry reduced effective loop models within the real framework, versus the self dual one \cite{CovSDSS, CovSDCosmo}.
%
Actually, the path from LQG to LQC is not straight. Despite recent progress \cite{Alesci:2016gub,Alesci:2015nja,Bodendorfer:2015qie,Livine:2011up, LQCAmbiguity},
LQC has not yet been defined as a cosmological sector of the full quantum theory in LQG. It is constructed as a LQG-inspired polymer quantization of the symmetry-reduced classical phase space, combined with a coarse-graining argument (to motivate the $\bar{\mu}$-scheme over the initial $\mu_{0}$-scheme). In this context, the BI parameter $\gamma$ appears in various effective parameters in LQC, the volume gap, the matter density at bounce, and so on. Although these effective parameters all descend from the unique fundamental BI parameter, the LQC phenomenology can vary them and adjust them independently due the flexibility of the derivation of LQC from LQG.
Here we propose to investigate precisely the canonical transformations generating the BU parameter, by properly defining the Thiemann complexifier and analyzing its flow on the cosmological phase space.

Our starting point is the CVH algebra of canonical general relativity, formed by the complexifier $\cC$, the 3d volume $v$ and the Hamiltonian constraint $\cH$, as identified by Thiemann \cite{ThiemannRC,Thiemann:1996aw} (see details in appendix \ref{CVHgen}). We consider it as a fundamental algebra underlying loop quantum gravity. It intertwines the scale transformation generated by the complexifier $\cC$ and the dynamics generated by the Hamiltonian constraint $\cH$. Our goal is to implement this algebra in LQC. In particular, we will show that, subtly adjusting the LQC-regularization, the CVH algebra remains closed and forms a $\su(1,1)$ algebra. First, this allows to exponentiate exactly the flow of both the Hamiltonian constraint and of the complexifier on the cosmological phase space. Then, it allows for a group theoretic quantization, along the lines sketched in \cite{LivGQCosmo}, in terms of $\SU(1,1)$ coherent states, where the regularized complexifier implements shifts in the BI parameter as unitary transformations in the quantum theory.

We believe that it is crucial to respect as much as possible the algebraic structure of observables during the quantization. From this perspective, it seems essential to keep a closed CVH algebra at the quantum level, especially since the Hamiltonian constraint and the complexifier are fundamental bricks of the LQG framework. Breaking this algebra would probably signal deeper quantization anomalies. In the present finite-dimensional setting of homogeneous cosmology, there is absolutely no reason to break this $\SU(1,1)$ algebra at the quantum level. In the more general setting of full LQG (or midi-superspace LQG models with inhomogeneities), the fate of the CVH algebra is likely related to the possibility of conformal anomalies.

\medskip

The paper is organized as follows. The first section starts with a review of the Hamiltonian analysis of the classical FRW cosmology for gravity coupled to a massless scalar field. We define the Thiemann complexifier and exponentiate it to rescaling transformations. We show the CVH algebra forms a closed $\su(1,1)$ Lie algebra and we compute its Casimir in terms of the matter density.
The second section introduces the regularization scheme of loop quantum cosmology, introducing a fundamental volume scale $\lambda$. Applying the  LQC regularization scheme to the complexifier, we show that it generates canonical transformations that rescale the Hamiltonian constraint without modifying the volume regularization scale. It is rather surprising to get such a simple definition of  a dilatation operator compatible with the scale truncation of the phase space. Further similarly regularizing the volume,  the CVH algebra still closes and we compute the $\su(1,1)$ Casimir in terms of the matter density and volume scale $\lambda$. This leads us to  a corrected ansatz for the effective classical Hamiltonian for LQC.

The third section shows how to exponentiate the Hamiltonian constraint and solve the dynamics exploiting the $\SU(1,1)$ group structure. The Hamiltonian constraint is identified as a null generator while the complexifier is a pure boost. We derive the corrected Friedman equations corresponding to our fully regularized ansatz. Furthermore we show that the complexifier generically becomes the  deparametrized Hamiltonian when deparameterizing the theory in terms of the scalar field time. This hints towards a deeper relation between the evolution for (quantum) gravity and scale transformations.

The fourth section is devoted to the quantization. The phase space is quantized in terms of $\SU(1,1)$ representations and the complexifier operator $\hat{\cC}$ generates unitary transformations $W$. An important feature is that the volume gap is now independent from the Barbero-Immirzi parameter. Shifting the BI parameter is more subtle than a straightforward rescaling of the geometric observables:
the initial volume operator $\hat{v}$ and its Immirzi-rescaled version $\hat{v}' =W \hat{v} W^{-1}$ do not commute and thus do not share the same eigenvectors basis. This situation is similar to  the area operator and boosted area operator not commuting with each other in LQG \cite{Rovelli:2002vp}. It allows to implement Immirzi boosts unitarily without changing Hilbert space and begs the question of how the CVH algebra extends to  full LQG and spin network states.

%%%%%%%%%%%%%%
\section{The classical setting: Standard FRW Cosmology}
%%%%%%%%%%%%%%

We place ourselves in the setting of the flat FLRW cosmological model ($k=0$), that is we restrict general relativity to a spatially flat, homogenous and isotropic spacetime. We consider the corresponding symmetry reduced gravitational phase space coupled to a massless scalar field.

%%%%%
\subsection{Classical phase space for homogeneous and isotropic cosmology}
%%%%%

We consider gravity coupled to a massless scalar field. Considering homogenous and isotropic fields, the Hamiltonian formulation of the coupled system simplifies drastically. The canonical variables are given by the scalar factor $a$ and its momentum $\pi_a$, as well as by the scalar field $\phi$ and its momentum $\pi_{\phi}$. This defines the phase space structure, provided with a Hamiltonian constraint corresponding to the Einstein equations applied to this symmetry-reduced setting:
\be
\{ a, \pi_a\} = 1\,,
\qquad 
\{ \phi, \pi_{\phi}\}= 1\,,
\qquad 
\cH^{0} = \frac{\pi^2_{\phi}}{2 a^3}- \frac{2 \pi G}{3}  \frac{\pi^2_a}{a}  = 0\,.
\ee
The superscript $0$ underlines that we are dealing with the standard FRW cosmology Hamiltonian and not yet the loop cosmology modified Hamiltonian as we will see in the next section.
This phase space can then be written in terms of the homogenous and isotropic Ashtekar-Barbero connection $A^i_a = c \delta^i_a$ and its conjugated momentum $E^a_i = p \delta^a_i$ where we have the following relation between the two kind of cosmological variables
\be
a = \sqrt{p}\,,
\qquad
\pi_a = - \frac{3}{4 \pi G \gamma} \sqrt{p} c\,.
\ee
Note that since we are working within the flat FLRW framework, the spin-connection vanishes and only the extrinsic curvature term survives in the Ashtekar-Barbero connection. Its expression reduces to $A = \Gamma + \gamma K = \gamma K = c = \gamma \dot{a}$. In this specific framework, the Ashtekar-Barbero connection is thus directly proportional to the BI parameter $\gamma$. 
Using the new set of canonical variables $(c,p)$ and $(\phi, \pi_{\phi})$, the phase space becomes
\be
\{ c, p \} = \frac{8 \pi G \gamma}{3}\,,
\qquad
\{ \phi, \pi_{\phi}\}= 1
\,,
\qquad
\cH^{0} = \frac{1}{16\pi G} \; \bigg{[} \; 8 \pi G \frac{\pi^2_{\phi}}{p^{3/2}} - \frac{6}{\gamma^2} \sqrt{p} c^2 \; \bigg{]} = 0\,.
\ee
Finally, we can introduce yet another set of canonical variables, which is most convenient to formulate loop quantum cosmology (LQC). We introduce the 3d volume $v$ and its conjugate momentum $b$. This new pair of canonical variables $(b,v)$ further simplifies the phase space and absorbs the numerical pre-factors in the Poisson bracket. The change of variable between $(c,p)$ and $(b,v)$ is given by:
\be
\label{(b,v)}
p = (4 \pi G v)^{2/3}
\,, \qquad
c = \gamma (4 \pi G v)^{1/3} b\,,
%\qquad
\ee
\be
4\pi G v = p^{3/2} = a^{3}
\,,\qquad
b = \gamma^{-1}c p^{-1/2}= -\f{4\pi G}3 a^{-2}\pi_{a}\,.
\ee
%\be
%4\pi G v = a^{3}\,, \quad b = -\f{4\pi G}3 a^{-2}\pi_{a}\,.
%\ee
%
Using these new canonical variables, the homogeneous cosmology phase space takes the final following form:
\be
\{ b, v\} = 1 \qquad \{ \phi, \pi_{\phi}\} = 1 \qquad \qquad \cH^{0} = \frac{\pi^2_{\phi}}{8 \pi G v} - \frac{3}{2} v b^2 = 0
\ee
%
%Note that because of the way we define the new canonical variables $(v,b)$ in (\ref{(b,v)}), the Immirzi parameter disappears from the phase space, both from the Poisson bracket and the constraint.
%
The BI parameter $\gamma$ appears in the phase space and Hamiltonian constraint only when we use the loop gravity connection-triad variables $(c,p)$. It disappears from the kinematics and dynamics of the theory if we work in the original scale factor variables $(a,\pi_{a})$ or volume variables $(v,b)$. This makes explicit that the BI parameter does not appear in the physical predictions of the theory at the classical level as expected.

%%%%%
\subsection{Cosmological dynamics from CVH algebra: the underlying $\sl_{2}$ structure}
%\subsection{The cosmological dynamics: the CVH algebra}
%%%%%

For our purpose, we rescale the Hamiltonian constraint by a $\f13$-factor and distinguish its gravitational component from its matter part:
\be
\label{Classical}
\cH^{0}=\cH^{0}_{g}+\cH^{0}_{m}
=
-\f12vb^{2}
+
\f{\pi_{\phi}^{2}}{24\pi G\,v}
%=-\f12vb^{2}+\f\beta2 v^{-1}\,,
%\qquad
%\beta= \f{\pi_{\phi}^{2}}{12\pi G}\,,
\ee
where the matter field energy-momentum contains an inverse volume factor.

Let us first focus on the gravitational part. Following the definitions of the full theory,  the Thiemann complexifier can also be introduced as the integral over a space-like region $\Sigma$ of the trace of the extrinsic curvature $K$. Its symmetry reduction to flat FLRW geometry leads to
\be
C \equiv \frac{1}{4\pi G \gamma}  \int_{\Sigma} d^3x \; E^a_i K^i_a =  \frac{V_0}{4\pi G \gamma}  \; p c = V_0\,  vb\,,
\ee
where $V_{0}$ is the volume of a fiducial space cell. Considering a unit fiducial volume $V_0 =1$, we obtain $C = vb$.
This means that the complexifier is the dilatation in the $(b,v)$ phase  space, exactly the same way that it is the dilatation in the $(K^i_a ,E^a_i)$ phase space for the full theory.

Further following the structures developed for full loop gravity,  the complexifier can also be derived as the Poisson bracket of the volume $v$ with the Hamiltonian constraint:
\be
\label{C}
C=\{v,\cH^{0}_{g}\}=vb\,.
\ee
Reversing the logic, this relation can be actually considered as the definition of the Thiemann complexifier $C$ and it is the point of view that we will take when turning to loop cosmology in section \ref{C_LQC}.
Moreover, the three observables $C$, $v$ and $\cH^{0}_g$ actually form a closed Lie algebra:
\be
\label{VH}
\{C,v\}=v
\,,\quad
\{C,\cH^{0}_{g}\}=-\cH^{0}_{g}
\,.
\ee
This is the gravitational CVH algebra in the cosmological setting, translating the CVH algebra of the full theory \cite{ThiemannRC} (see appendix A for more details) to the highly symmetric context of FRW cosmology.
We recognizes it as a $\sl_{2}$ algebra with vanishing  Casimir:
\be
\label{CazClass}
\mathfrak{C}_{\sl_{2}}= -  2v\cH^{0}_{g} - C^{2}  =0\,.
\ee
We can re-organize the generators to set the algebra in its usual presentation. We define the linear combinations:
\be
C=k_{y}\,,\quad
v=\f12(j_{z}+k_{x})\,,\quad
\cH^{0}_{g}=k_{x}-j_{z}\,,
\ee
\be
j_{z}=v-\f12\cH^{0}_{g}
=v\left(1+\f{b^{2}}4\right)\,,\quad
k_{x}=v+\f12\cH^{0}_{g}
=v\left(1-\f{b^{2}}4\right)\,,\quad
k_{y}=C=vb\,,
\ee
which leads to the standard form of the $\su(1,1)$ Lie algebra (with correct reality conditions), with a two boost generators and a spatial rotation:
\be
\label{CommutationRelations}
\{j_{z},k_{x}\}=k_{y}\,,\quad
\{j_{z},k_{y}\}=-k_{x}\,,\quad
\{k_{x},k_{y}\}=-j_{z}\,,\quad
\mathfrak{C}_{\sl_{2}}=j_{z}^{2} - k_{x}^{2} - k_{y}^{2}=0
\,.
\ee
%\be
%C=K_{2}\,,\quad
%v=\f12(J_{3}+K_{1})\,,\quad
%\cH^{0}_{g}=K_{1}-J_{3}\,,
%\ee
%\be
%J_{3}=v-\f12\cH^{0}_{g}
%=v\left(1+\f{b^{2}}4\right)\,,\quad
%K_{1}=v+\f12\cH^{0}_{g}
%=v\left(1-\f{b^{2}}4\right)\,,\quad
%K_{2}=C=vb\,,
%\ee
%\be
%\{J_{3},K_{1}\}=K_{2}\,,\quad
%\{J_{3},K_{2}\}=-K_{1}\,,\quad
%\{K_{1},K_{2}\}=-J_{3}\,,\quad
%\mathfrak{C}_{\sl_{2}}=K_{1}^{2}+K_{2}^{2}-J_{3}^{2}=0
%\,.
%\ee
%
The vanishing Casimir means that the system will be represented at the quantum level by a null representation of $\SU(1,1)$.

A crucial remark is that we can extend this CVH algebra to the full Hamiltonian constraint and take into account the matter field contribution:
\be
C_{full}\equiv\{v,\cH^{0}\}=vb=C\,,\quad
\{C,\cH^{0}\}=-\cH^{0}
\,,\quad
\{C,v\}=v
\,.
\ee
This allows to describe the whole system gravity plus matter field at the dynamical level through the same $\SU(1,1)$ structure. We define similarly the $\su(1,1)$ generators, using capital letters to distinguish them from the definitions of the pure gravitation sector:
\be
J_{z}=v-\f12\cH^0
\,,\quad
K_{x}=v+\f12\cH^0
\,,\quad
K_{y}=C=vb\,.
\ee
This shows that the evolution flow generated by the Hamiltonian constraint can be exactly integrated as a Lorentz transformation.
What changes compared to the pure gravitation sector is that the matter term affects the value of the Casimir:
\be
\label{CazClassMat}
\mathfrak{C}_{\sl_{2}}= - 2v\cH^{0}-C^{2}= -  \f{\pi_{\phi}^{2}}{12\pi G} < 0\,,
\ee
where $\pi_{\phi}$ is a constant of motion. This means that due to the matter energy term the system will  be represented at the quantum level by a space-like representation of $\SU(1,1)$.
This is slightly different from the group quantization framework for loop quantum cosmology introduced in \cite{LivGQCosmo} where the deparameterized dynamics of the quantum space-time with respect to the scalar field clock was described in terms of time-like $\SU(1,1)$-representations.

%%%%%
\subsection{Immirzi transformations and Geometry Rescaling}
%%%%%

As in the full theory, we expect the observable $C$ to play the role of the {\it Thiemann complexifier}, that is, generate the canonical transformations shifting the value of the BI parameter.
First of all, the BI parameter does not appear when using the $(v,b)$ variables, so the the Hamiltonian constraint should be invariant under the flow generated by $C$. On the other hand, if we switch back to the $(c,p)$ variables, we should see how transformations generated by $C$ modifies the canonical bracket and the Hamiltonian constraint.

Let us start with the $(v,b)$ variables. The complexifier $C$ is simply the dilatation $vb$ and it is direct to exponentiate its action:
\be
\label{Cflow}
v\longrightarrow
\tilde{v}=
e^{\eta \{C,\cdot\}}\, v =e^{\eta}v
\,,
\qquad
b\longrightarrow
\tilde{b}=
e^{\eta \{C,\cdot\}}\, b =e^{-\eta}b\,.
\ee
Both matter and gravitational parts of the Hamiltonian constraint transform homogeneously under these transformations:
\be
\cH^{0}\longrightarrow
\tilde{\cH}^{0}=
e^{\eta \{C,\cdot\}}\, \cH^{0} =e^{-\eta}\cH^{0}\,.
\ee
Such rescaling clearly does not affect the constraint $\cH^{0}=0$ and the theory is as expected invariant under dual rescaling of the volume $v$ and its conjugate variable $b$.

%\smallskip

Let us check how the complexifier works on the Ashtekar-Barbero variables $(c,p)$. The complexifier $C$ also generates a straightforward dilatation of the variables $c$ and $p$:
\be
C=vb=\f1{4\pi G \gamma} cp\,,
\qquad
\begin{array}{l}
p\rightarrow
\tilde{p}=
e^{\eta \{C,\cdot\}}\, p =e^{\f23\eta}p
\\
c\rightarrow
\tilde{c}=
e^{\eta \{C,\cdot\}}\, c =e^{-\f23\eta}c
\end{array}
\ee
The usual way to proceed in loop gravity is to keep the same triad variable $p$ and change only the connection variable, $c$ to $\tilde{c}$. This leads to a modified Poisson bracket:
\be
\{ c, p \} = \frac{8 \pi G \gamma}{3} 
\,\longrightarrow\,
\{ \tilde{c}, p \} = \frac{8 \pi G \tgamma}{3}  
\quad\textrm{with}\,\,
\tgamma=e^{-\f23\eta}\gamma\,,
\ee
with a rescaled BI parameter $\tgamma$. This modification of the kinematical phase space structure leads to a similar transformation of the Hamiltonian constraint:
\begin{align}
&\cH^{0}_{\gamma}[c,p] =
 \frac{\pi^2_{\phi}}{2p^{3/2}}
 -\frac{3}{8\pi G\gamma^2} \sqrt{p} c^2 \\
&\longrightarrow
\quad \tilde{\cH}^{0}[\tc,p]=
e^{\eta \{C,\cdot\}}\,\cH^{0} =
 \frac{\pi^2_{\phi}}{2\tp^{3/2}}
 -\frac{3}{8\pi G\gamma^2} \sqrt{\tp} \tc^2
 =
e^{-\eta}
\left(  \frac{\pi^2_{\phi}}{2p^{3/2}}
 -\frac{3}{8\pi G\tgamma^2} \sqrt{p} \tc^2\right)
 =
e^{-\eta} \cH^{0}_{\tgamma}[\tc,p]
\,.\nn
\end{align}
%\be
%\cH^{0}_{\gamma}[c,p] =
% \frac{\pi^2_{\phi}}{2p^{3/2}}
% -\frac{3}{8\pi G\gamma^2} \sqrt{p} c^2
%\longrightarrow
%\tilde{\cH}^{0}[\tc,p]=
%e^{\eta \{C,\cdot\}}\,\cH^{0} =
% \frac{\pi^2_{\phi}}{2\tp^{3/2}}
% -\frac{3}{8\pi G\gamma^2} \sqrt{\tp} \tc^2
% =
%e^{-\eta}
%\left(  \frac{\pi^2_{\phi}}{2p^{3/2}}
% -\frac{3}{8\pi G\tgamma^2} \sqrt{p} \tc^2\right)
% =
%e^{-\eta} \cH^{0}_{\tgamma}[\tc,p]
%\,.
%\ee
%
This shows how Thiemann complexifier, given by the dilatation on the volume phase space $(b,v)$, does indeed generate shifts in the BI parameter as claimed.

%%%%%%%%%%%%%%
\section{The Immirzi generator for  Loop Cosmology}
\label{C_LQC}
%%%%%%%%%%%%%%

%%%%%
\subsection{Regularized  Hamiltonian for Loop Quantum Cosmology}
%\subsection{Effective Hamiltonian for Loop Quantum Cosmology}
%\subsection{Deformed Hamiltonian for Loop Quantum Cosmology}
%%%%%

We would like to start by studying the Thiemann complexifier for loop quantum cosmology and the fate of the BI parameter at the classical level. For this purpose, we consider the effective classical framework for LQC. This can be seen for two dual, but ultimately equivalent perspectives:

\begin{itemize}

\item One starts with LQC, as the polymer quantization of classical  cosmology obtained from a symmetry-reduction of classical general relativity. This quantization procedure follows the same techniques as used for loop quantum gravity, but LQC can not be directly derived as a symmetry-reduction of LQG at the quantum level. We then compute the effective classical dynamics of peaked wave packets, i.e. coherent states, which shows how quantum gravity effects regularize the curvature observables for general relativity. 

\item Alternatively, again starting with classical cosmology, one can directly implement the regularization scheme at the classical level and define a deformed classical cosmology theory, which can then be straightforwardly canonically quantized into LQC.

\end{itemize}

Both ways to proceed lead to loop quantum cosmology. The crucial point is the regularization of the observables entering the Hamiltonian constraint algebra, which relies on replacing the curvature $F[A]$ by the holonomy $U[A]$ around a tiny but finite loop, as in lattice gauge theory.
%of size given by regularization scale $\lambda$. 
The difference between the two viewpoints is whether this regularization scheme is a consequence of
the quantization procedure or a prerequisite that allows for the loop quantization.

\smallskip

So the fundamental idea underlying LQC is to quantize the theory in a different representation than the usual Schr\"odinger one and  instead base the quantum theory on the now-called polymer representation. We do not consider the observable $b$ on the classical phase space but its exponentiated version $e^{i \lambda b}$. At the quantum level, the operator $\hat{b}$ does not exist anymore, only the ``holonomy'' $\widehat{e^{i \lambda b} }$ is well-defined\footnotemark{}  on the Hilbert space.
\footnotetext{In particular, the operator $e^{i \lambda b}$ is not weakly continuous in $\lambda$, so the operator $\hat{b}$ can not be recovered from infinitesimal variations.
}
Only in the limit $\lambda \rightarrow 0$ one recovers the usual classical phase space out of the polymerized version.
In practice, this implies regularizing\footnotemark{} the curvature $F[A] \propto v b^2$ using a suitable  exponentiated version of $b$.
\footnotetext{
This procedure is  not unique and leads to regularization ambiguities, which can be understood as using superpositions of harmonics of the basic regularization scale (see \cite{LQCAmbiguity} for a recent study of these ambiguities and their consequences in LQC). However, the common choice the LQC literature is to regularize the observable $b$ using the fundamental mode $e^{i \lambda b}$ (see \cite{LQCReport} for details).
}
The standard polymer regularization proceeds to replace the observable $b$ by the simplest function of the ``holonomy'' $e^{i \lambda b}$ admitting the correct limit when the regularization scale $\lambda$ is sent to 0:
\be
b \rightarrow \frac{\sin{(\lambda b)}}{\lambda} = \frac{ e^{i \lambda b} - e^{- i \lambda b}}{2 i\lambda}  \,.
\ee
The polymerized cosmological phase space becomes:
%After polymerization, the homogenous and isotropic phase space of gravity becomes
\be
\{ e^{i \lambda b}, v\} =  i \lambda e^{i \lambda b}\,, \quad \{ \phi, \pi_{\phi}\} = 1\,,
\qquad 
\cH=\cH_{m}+\cH_{g}
= \frac{\pi^2_{\phi}}{24 \pi G v} - \frac{1}{2} v \frac{\sin^2{(\lambda b)}}{\lambda^2} = 0\,,
\label{effH}
\ee
where we have introduced the effective regularized Hamiltonian constraint $\cH$. 
We have already rescaled the Hamiltonian constraint by a $\f13$-factor to fit with our choice of normalization for the standard FRW cosmology case.
This standard procedure regularizes only the gravitational  Hamiltonian $\cH_{g}$ and does not a priori affect the matter part $\cH_{m}$.

\smallskip

Let us underline  that, in this classical effective framework, the Barbero-Immirzi parameter $\gamma$ does not appear, neither in the phase space nor in the regularized Hamiltonian constraint \footnotemark.
\footnotetext{  Note that this is a special property of the homogeneous and isotropic Ashtekar-Barbero phase space. In the full theory, while one can always rescale the electric field to remove $\gamma$ from the canonical bracket, there is no way to remove the BI parameter from the Hamiltonian scalar constraint unless fixing it to some specific value. In our symmetry reduced phase space, because of the  Levi-Civita connection vanishes, a simple redefinition of the canonical variables allows to remove totally the BI parameter from the phase space, without fixing $\gamma$ to any specific value.}
What matters is the regularization scale $\lambda$ and not the BI parameter. If we do not add any input from the full LQG theory and we simply quantize this phase space using the polymer representation, there will be no trace of the BI parameter in the resulting quantum theory.
The BI parameter appears in LQC when we think of it as an effective model descending from the fundamental LQG theory. Then, assuming as in standard LQC that the regularization scale $\lambda$ is related to the kinematical area gap derived in full LQG, the BI parameter $\gamma$ reenters in the model since it controls the kinematical area gap (and volume gap): \be
\label{lg}
\lambda^2_{eff}= 4\sqrt{3} \pi \gamma l^2_{\text{Pl}}\,.
\ee
Plugging this relation in the regularized Hamiltonian constraint re-introduces the BI parameter in the theory.

Here we will develop an alternative point of view for LQC and keep the distinction between the BI parameter $\gamma$ and the regularization scale $\lambda$. Indeed, we will define below a regularized complexifier, which turns out to generate shift in the BI parameter while keeping the regularization scale  $\lambda$ fixed. We still get a closed CVH algebra. This will lead us to a group quantization of the regularized phase space, where  the existence of a  volume gap is encoded through the choice of representation. The complexifier then generates unitary transformations in the quantum theory without changing the volume spectrum: the BI parameter (as generated by the complexifier) disappears from the physical predictions while the regularization scale keeps its essential role. This underlines the importance of properly defining what is meant by ``BI parameter'' and of not confusing its multiple facets, as a parameter labeling canonical transformations on the classical phase space or as a physical parameter determining the fundamental size of the quanta of geometry. This confusion leads in LQG and its related symmetry reduced polymer models to the so called Immirzi ambiguity. This ambiguity takes its origin in the fact that the canonical transformation introduced by Barbero, and generated by the complexifier, is not mapped to a unitary transformation in the quantum theory, as one could expect. Because of this unnatural feature of the current loop quantization, the kinematical and physical predictions of the related quantum models always depend on the BI parameter, which does not play any role in the classical theory.

In the next sections, we will show how the Barbero canonical transformation can be mapped to a unitary transformation in the quantum theory, by extending the regularization scheme of standard LQC with the requirement to preserve the CVH algebra. As we shall see, in this new LQC inspired model, the BI parameter $\gamma$ and the regularization scale $\lambda$ are totally disantangle, and the Immirzi ambiguity disappears. While $\gamma$ is rescaled under the action of the complexifier, the scale $\lambda$ remains unaffected under the same action. This is one of the main results of our construction, along with a new exactly solvable model for LQC and interesting insights on the role of the complexifier as the generator of the deparametrized dynamic.
 In future investigation, it will be interesting to see if those new features can be extended to the full loop quantum gravity theory.

% as we will show in sections \ref{fullreg1} and \ref{fullreg2}.

%%%%%
\subsection{The gravitational CVH algebra: Immirzi transformations as boosts}
%%%%%

Now that we have defined the regularized phase space and Hamiltonian constraint, we investigate the action of the complexifier.
One point of special interest to us is the fate of CVH algebra at the effective level, i.e. after having introduce the loop regularization, or, in other words, what are the transformations of the volume and scalar constraint under the action of the complexifier.

Let us start by considering the action of undeformed complexifier $C=vb$ on the Hamiltonian constraint $\cH_{\lambda}$. It rescales  the volume $v$ and its conjugate angle $b$, which can be re-interpreted as a rescaling of the regularization parameter $\lambda$:
\be
\cH_{\lambda} = \frac{\pi^2_{\phi}}{24 \pi G v} - \frac{1}{2} v \frac{\sin^2{(\lambda b)}}{\lambda^2}
\,\longrightarrow\,
\tilde{\cH}_{\lambda}[\tv,\tb]
= \frac{\pi^2_{\phi}}{24 \pi G \tv} - \frac{1}{2} \tv \frac{\sin^2{(\lambda \tb)}}{\lambda^2}
={\cH}_{\lambda}[\tv,\tb]
= \cH_{e^{-\eta}\lambda}[v,b]\,.
\ee
Assuming that $\lambda$ is directly proportional to $\gamma$, this would be interpreted as a rescaling of the BI parameter, which fits perfectly with the assumed relation (\ref{lg}). 

\textit{However, from the point of view of the loop (polymer) quantization, working with the classical expression of the complexifier $C = vb$ is not well defined.} Indeed, when going to the quantum theory, the operator $\hat{b}$ is ill-defined in the framework of loop quantum cosmology. Only its exponentiated version $\hat{e^{i\lambda b}}$  has a well defined action. It seems therefore necessary to also regularize the complexifier $C$. However, this regularization should not be introduced arbitrarily. In this work, we insist on regularizing the complexifier such that we preserve the structure of the classical CVH algebra. This leads to a new canonical transformation of the regularized phase space (see Section C) which do not change the regularization scale $\lambda$, but nevertheless rescale the volume. In the classical limit, ie when $\lambda \rightarrow 0$, our regularized complexifier concides with the classical expression $C =vb$ and give back the standard Immirzi generator. This constitute our new proposal for the implementation of the regularized Barbero canonical transformation in loop quantum cosmology. 

Let us now compute our regularized complexifier. The requirement of preserving the CVH algebra structure implies that we should use, at the effective level, the classical definition \eqref{C} of the complexifier as the Poisson bracket of the gravitational part of the Hamiltonian constraint and the volume. It reads
\be
\label{newcom}
\cC\equiv\{v,\cH_{g}\}=v\f{\sin 2\lambda b}{2\lambda} \quad\underset{\lambda\sim0}\sim vb\,.
\ee
Since $\cC$ reduces to the classical generator of the Barbero canonical transformation, ie the classical complexifier, when the regularization scale $\lambda$ is sent back to 0, it can be legitimately thought as the suitable regularization of the complexifier. Yet it does not generate anymore a canonical transformation on the classical canonical variables $(b,v)$, but on a new set of regularized canonical variables $(B,V)$ that we introduce in the next section, Eq (\ref{NVC1}-\ref{NVC2}). When $\lambda \rightarrow 0$, the regularized version of the canonical transformation generated by our new complexifier (\ref{newcom}) on $(B, V)$ reduces to the one generated by the classical complexifier $C =vb$ on the initial classical canonical variables $(v,b)$. In this sense, we have obtain a generalized Barbero canonical transformation with the right classical limit. The problem of the Immrizi ambiguity in LQG, as we undertsand it, is that the Barbero canonical transformation is not mapped to a unitary transformation in the quantum theory. In this work, we will show that the generalized Barbero canonical transformation on the canonical variables $(B,V)$ can be mapped to a unitary transformation through our quantization procedure. In this precise sense, we claim that the Immirzi ambiguity is resolved by our quantization procedure based on the CVH algebra. 

Computing the two other brackets of the CVH algebra, we obtain a closed system
\be
\{\cC,v\}=v+4\lambda^{2}\cH_{g}\,,\quad
\{\cC,\cH_{g}\}=-\cH_{g}\,,
\ee
with a slight correction compared to the classical FRW case. So the $\sl_{2}$ algebraic structure is preserved by the loop regularization. We  introduce the following $\sl_{2}$ generators
\be
j_{z}=(2\lambda)^{-1}v\,,\quad
k_{\pm}=(2\lambda)^{-1}ve^{\pm 2\lambda i b }\,,\quad
\{j_{z},k_{\pm}\}=\mp i k_{\pm}\,,\quad
\{k_{+},k_{-}\}=2i j_{z}\,.
\ee
the volume, complexifier and scalar constraint being given by
\be
v=2\lambda j_{z}\,,\quad
\cC=k_{y}=\f1{2i}(k_{+}-k_{-})\,,\quad
\cH_{g}=(2\lambda)^{-1}(k_{x}-j_{z})
\ee
Since $k_{+}=\overline{k_{-}}$, this algebra satisfies the reality conditions of a $\su(1,1)$ Lie algebra. Its Casimir vanishes:
\be
\label{CazLQC}
\mathfrak{C}_{\sl_{2}}
=j_{z}^{2} - k_{+}k_{-}
=j_{z}^{2} - k_{x}^{2}+k_{y}^{2}
=0\,.
\ee
The quantum theory will naturally shift this vanishing Casimir to a positive Casimir, similarly to the ground state energy of the quantum harmonic oscillator. This non-vanishing Casimir at the quantum level corresponds to a volume gap and defines the lowest possible volume state. It is also possible to take into account this volume gap at the classical level and introduce by hand a minimal volume $v_{m}$ in the definition of the $\su(1,1)$ generators. This actually defines the most general $\su(1,1)$ phase space in our framework, as explained in detail in the appendix \ref{vmin}.
%
%{\Jib From the point of view of the quantum theory, this can be too restrictive. In Appendix, we show how one can extend the formalism to a positive Casimir, by introducing a new step in the regularization involving a modification of the volume and therefore, a new scale $v_m$. See Appendix for more details.}
 
Moreover, we see that the new complexifier $\cC$ is also regularized, with $b$ turned into a periodic trigonometric function $(2\lambda)^{-1}\sin 2\lambda b$. Moreover we can use the $\SU(1,1)$ group structure to integrate the action of the regularized complexifier $\cC$ as boost transformations.
It is straightforward to exponentiate the action of the $\su(1,1)$ generators $\vj=(j_{z},k_{x},k_{y})$. The most direct method is to introduce the 2$\times$2 Hermitian matrix:
\be
M=\mat{cc}{j_{z}&k_{-}\\k_{+}&j_{z}}\in\,H_{2}(\C)\,.
\ee
for which the determinant give the $\su(1,1)$ Casimir. Computing the Poisson bracket of the generators $\vj=(j_{z},k_{x},k_{y})$ with this matrix $M$, which is simply a re-writing of the $\sl_{2}$ Poisson algebra, we obtain
\be
\{\vJ,M\}\,=\,\f i2\,\left(\vtau\,M\,-\,M\,\vtau^{\dagger}\right) 
\ee
where the $\tau_{a}$ are the Lorentzian Pauli matrices:
\be
\tau_{z}= \frac{1}{2} \mat{cc}{1 & 0 \\ 0 & -1}= \frac{1}{2} \sigma_{z},
\quad
\tau_{x}= \frac{1}{2} \mat{cc}{0 & 1 \\ -1 & 0}=+\frac{i}{2} \sigma_{x},
\quad
\tau_{y}= \frac{1}{2} \mat{cc}{0 & -i \\ -i & 0}=-\frac{i}{2}\sigma_{y}\,.
\ee
In order to recover the $\su(1,1)$ commutation relations (\ref{CommutationRelations}), one set $k_x = i \tau_x$, $k_y = i \tau_y$ and $j_z = i \tau_z$.
This Poisson bracket can be exponentiated into finite $\SU(1,1)$ transformations :
\be
M\,\longrightarrow\, \tilde{M}=e^{\{\veta\cdot\vJ,\,\cdot\,\}}\,M=G\,M\,G^{\dagger}\,,
\ee
\be%\qquad
\textrm{with}\quad
G=e^{i\,\veta\cdot\vtau}
=\mat{cc}{\alpha & \beta \\ \bbeta & \balpha}\in\SU(1,1),
\quad
|\alpha|^2-|\beta|^2=1\,. \nn
\ee 
Another more systematic way to proceed is to use the spinorial parametrization of the phase space as explained in \cite{LivGQCosmo} and reviewed in appendix B. This spinorial formalism is especially useful to proceed to the coherent state quantization of the theory.

Now, we focus on the flow generated by the complexifier $\cC=k_{y}$. The corresponding $\SU(1,1)$ transformation is simply\footnotemark{}:
\be
\label{Cregflow}
e^{\eta \{\cC,\cdot\}} M= G_{\eta}MG_{\eta}^{\dagger}
\,,\quad
G_{\eta}
= \mat{cc}{\cosh{\f\eta2} & \sinh{\f\eta2} \\ { \sinh}{\f\eta2} & {\cosh}{\f\eta2} }
%= \mat{cc}{\cosh{(\eta/2)} & \sinh{(\eta / 2)} \\ { \sinh}{(\eta / 2)} & {\cosh}{(\eta/2)} }
=G_{\eta}^{\dagger}
\,.
\ee
\footnotetext{
Having in mind a possible Wick-rotation of the theory from a real BI parameter back to the original self-dual formulation theory with $\gamma=\pm i$, there is actually absolutely no problem  taking a complex transformation parameter $\eta$.  Then, for $\eta\in\C$, the transformation matrix $G_{\eta}$ does not necessarily lay in the unitary group $\SU(1,1)$, but in its complexification $\SU(1,1)_{\C}=\SL(2,\C)$.
For instance, in the purely imaginary case $\eta=\pm i \f\pi 2$, we have:
\be
\cosh\f\eta 2=\f1{\sqrt2}
\,,\quad
\sinh\f\eta 2=\pm\f i{\sqrt2}
\,,\quad
\cosh\eta =0
\,,\quad
\sinh\eta =\pm i
\,,
\quad
%\nn
%\ee
%\be
G_{\pm i \f\pi 2}=\f1{\sqrt 2}\mat{cc}{1 &i \\ i& 1}
\,,\quad
\det G_{\pm i \f\pi 2}=1
\,.
\nn
\ee
The corresponding rescaling of the volume and the gravitational Hamiltonian is also complex:
\be
\cH_{g}\rightarrow \widetilde{\cH}_{g}=\mp i \,\cH_{g}
\,,\qquad
v\rightarrow \tv =\pm i \,(v + 4\lambda^{2}  \cH_{g})\,.
\nn
\ee
Note that even though the Hamiltonian constraint is now purely imaginary, this is only due to the numerical pre-factor and it actually still defines the exact same constraint.}
Extracting from this formula the transformations for the volume $v$ and the gravitational component of the Hamiltonian constraint, we get:
\be
\label{TransfoVH}
\begin{array}{lcl}
\;\;\; v
=(2\lambda)j_{z}
&\longrightarrow\,&
\tv
=(2\lambda)( j_{z}\cosh\eta+k_{x}\sinh\eta ) = e^{\eta}v +4\lambda^{2}\cH_{g}\sinh\eta
\\
\cH_{g}=(2\lambda)^{-1}(k_{x}-j_{z})
&\longrightarrow\,&
\widetilde{\cH}_{g}
=(2\lambda)^{-1}(\tilde{k}_{x}-\tilde{j}_{z})
=e^{-\eta}\,\cH_{g}\,.
\end{array}
\ee
First, the gravitational Hamiltonian constraint gets simply rescaled, as in the previous case for standard FRW cosmology. However, the regularization scale $\lambda$ does not get rescaled. Since the complexifier is supposed to generate rescaling of the BI parameter, this means that we should distinguish the regularization scale  $\lambda$ from the BI parameter $\gamma$.
Second, the volume is not simply rescaled. The action of the regularized complexifier is not a simple rescaling of the $(v,b)$ phase space, as before. Actually, the new volume observable does not commute anymore with the original volume:
\be
\{v,\tv\}
=4\lambda^{2}\sinh\eta\,\{j_{z},k_{x}\}
=4\lambda^{2}\sinh\eta\,\cC
=4\lambda^{2}\sinh\eta\,\{v,\cH_{g}\}\ne 0\,.
\ee
This feature will be carried to the quantum level: the complexifier boosts the volume operator and volume eigenstates change under the action of the complexifier.
This is reminiscent of the action of Lorentz boost (as generated by the Hamiltonian constraint) in loop quantum gravity on geometrical observables: the boosted area does not commute with the area in the original reference frame \cite{Rovelli:2002vp,LorentzLQG2}.

To summarize, we have a regularized complexifier $\cC$, which forms a closed Lie algebra under the Poisson bracket with the volume $v$ and the gravitational Hamiltonian $\cH_{g}$.
Since the loop quantization scheme compactifies in some sense the extrinsic curvature by considering the $\SU(2)$ holonomies of the Ashtekar-Barbero connection, it seems fair to also regularize the complexifier $\cC$, which is actually the integrated extrinsic curvature.
Then this regularized complexifier generates canonical transformations on the cosmological phase space, that do not change the regularization scale $\lambda$ but still rescales both the volume and the gravitational Hamiltonian.
It preserves the $\su(1,1)$ structure and thus allows to develop a group quantization of (the gravitational sector of) this effective loop cosmology phase space.
This is our proposal for Immirzi transformation in loop quantum cosmology. 

Finally, let us stress that the $\SU(1,1)$ transformations are linear transformations on the $\su(1,1)$ generators and do not affect the $\sl_{2}$ Casimir, which still vanishes in the present case. At the quantum level, this will be implemented as unitary transformations on the Hilbert space of the theory and will not require to change the $\SU(1,1)$ representation. In particular, Immirzi transformations will be unitary operators, which is a novel feature compared to previous LQC models sine it solves in this simplest framework the so called Immizi ambiguity.

%%%%%

%%%%%
\subsection{The full CVH algebra and Consistent Effective LQC Hamiltonian}
\label{fullreg2}
%%%%%

Up to now, we have focused on the complexifier acting on the gravitational Hamiltonian. In the case of standard FRW, as we showed in the first section, we could extend this action to the full Hamiltonian with both gravitational and matter parts.

Unfortunately, a simple glance of the Hamiltonian constraint shows that the matter part is in $v^{-1}$ and does not get simply rescaled under the action of the complexifier:
\be
\begin{array}{lcl}
v^{-1}
&\longrightarrow\,&
\tv^{-1}
= (e^{\eta}v +4\lambda^{2}\cH_{g}\sinh\eta)^{-1}
\\
\cH_{g}
&\longrightarrow\,&
\widetilde{\cH}_{g}
=e^{-\eta}\cH_{g}\,.
\end{array}
\nn
\ee
This can nevertheless be fixed by realizing that the inverse volume factor  should itself probably be  regularized and also acquire a $b$-dependent correction factor. While it can seems unusual to add holonomy correction to the volume, and moreover in the matter sector, it turns out that such dependence of the volume on the connection $b$ is also present in spherical symmetry (albeit in the gravitational sector), as discussed in details in \cite{BojowaldVol}. Moreover, in the covariant effective approach to scalar perturbations in LQC, a similar correction, in spirit, is required to preserve the covariance of the perturbed geometry. Indeed, in this framework, an additional holonomy correction is introduced in the second order contribution to the matter sector of the scalar constraint, providing a deformed notion of covariance for the system gravity plus matter \cite{CovSpher,CovGowdy}. In the present work, such holonomy corrections allow us to preserve, at the effective level, the exact structure of the classical CVH algebra \eqref{VH}, even when including matter.

The only way to get the same transformation for the gravitational Hamiltonian $\cH_{g}$ and the inverse volume term $v^{-1}$ is to realize that the combination $j_{z}-k_{x}$ gets rescaled by $e^{-\eta}$ while the opposite combination $j_{z}+k_{x}$ gets rescaled by the inverse factor $e^{\eta}$. Thus we introduce a modified, or regularized, volume:
\be
\label{NVC1}
V\equiv (2\lambda)^{-1}\,\f12(j_{z}+k_{x})=v\cos^{2}\lambda b
\quad
\underset{\lambda\sim0}\sim v\,.
\ee
It coincides with the usual volume observable as the regularization scale $\lambda$ is sent to 0. But it now has a simple behavior under the exponentiated action of the complexifier $\exp(\eta\{\cC,\cdot\})$:
\be
V\longrightarrow\tilde{V}=e^{+\eta}V\,.
\ee
This regularized volume is therefore the suitable generalization of the classical volume $v$ which preserves the action of the regularized complexifier. It should therefore corresponds to a new set of regularized canonical variables $(B, V)$. It is straiforward to obtain its canonically conjugated variable $B$ which reads
\be
\label{NVC2}
B = \frac{\tan{(\lambda b)}}{\lambda} \qquad \text{such that} \qquad \{ B, V\} = 1
\ee 
Under the exponentiated action of the complexifier $\exp(\eta\{\cC,\cdot\})$, it transforms as
\be
B\longrightarrow\tilde{B}=e^{-\eta}B\,. \qquad \text{hence} \qquad \{ \tilde{B}, \tilde{V}\}= 1
\ee
Consequently, the regularized complexifier generates indeed a canonical transformation on this canonical variables, acting by dilatation/contraction on each of them. This canonical transformation is a generalization of the classical Barbero canonical transformation (\ref{Cflow}) on the classical variables $(b,v)$, which shift the BI parameter. It reduces to (\ref{Cflow}) in the classical limit $\lambda \rightarrow 0$.

From this new viewpoint, we propose a modification of the effective Hamiltonian ansatz for loop quantum cosmology to:
\be
\cH^{reg}\equiv
%\frac{\pi^2_{\phi}}{24 \pi G v} - \frac{1}{2} v \frac{\sin^2{(\lambda b)}}{\lambda^2}
\frac{\pi^2_{\phi}}{24 \pi G V} - \frac{1}{2} v \frac{\sin^2{(\lambda b)}}{\lambda^2}
=
\frac{\pi^2_{\phi}}{24 \pi G v\cos^{2}\lambda b} - \frac{1}{2} v \frac{\sin^2{(\lambda b)}}{\lambda^2}\,,
\label{Hreg}
\ee
with a regularization of the inverse volume factor in the matter Hamiltonian.
Note that, although this new regularization can appear unusual at first, it is actually a simple rescaling of the scalar constraint, with a renormalization of the regularization scalar $\lambda \rightarrow 2\lambda$. Indeed, we can factor out $\cos^2{(\lambda b)}$ and write:
\be
\cH^{reg}
=
\f1{\cos^2{(\lambda b)}}
\,\Bigg{[}
\frac{\pi^2_{\phi}}{24 \pi G v} - \frac{1}{2} v \frac{\sin^2{(2\lambda b)}}{(2\lambda)^2}
\Bigg{]}
\nn
\ee
The pre-factor $\cos^2{(\lambda b)}$ can either be re-absorbed in the lapse, so as a change of time coordinate. It does not change the solution of the constraint $\cH^{reg}=0$ and thus does not affect the equations of motion of physical trajectories.
 Note also that the classical limit $\lambda \rightarrow 0$ is well-defined and one recovers the standard FRW Hamiltonian constraint and phase space in this limit of a vanishing regularization scale \footnote{Note that our regularization can be further generalized. In order to avoid divergencies at the bounce, the inverse volume term can be slightly modified as $V^{-1} \rightarrow (V + \alpha)^{-1}$ where $\alpha$ is some constant different from zero. It does not change the different brackets and allows to avoid divergencies appearing at maximal density in the inverse volume term of the hamiltonian due to the $\cos^{-2}{(\lambda b)}$. This step make our new regularization safe at the bounce.}.

With this extension, the full Hamiltonian constraint, with both matter and gravitational contributions, now form a closed CVH algebra:
\be
\{V,\cH^{reg}\}=\cC
\,,\quad
\{\cC,V\}=V
\,,\quad
\{\cC,\cH^{reg}\}=-\cH^{reg}\,,
\ee
where we use both regularized complexifier and volume. The complexifier is again given by
\be
\cC = v \frac{\sin{(2\lambda b)}}{2\lambda}
\ee
The Casimir does not vanish anymore:
\be
\label{CazLQCCMat}
\mathfrak{C}_{\sl_{2}}=- 2V\cH^{reg}-\cC^{2}= - \f{\pi_{\phi}^{2}}{12\pi G}\,,
\ee
and is entirely fixed by the matter energy density. This fits perfectly with the case of the standard FRW cosmology, where the matter energy density also fixes the value of the $\sl_{2}$ Casimir in \eqref{CazClassMat}.

Explicitly, the generators of $\SU(1,1)$ now read:
\be
\cC=K_{y}
\,,\quad
V=\lambda(K_{x}+J_{z})
\,,\quad
\cH^{reg}=(2\lambda)^{-1}(K_{x}-J_{z})
\,,
\nn
\ee
\be
K_{y}=\cC=v\f{\sin 2\lambda b}{2\lambda}
\,,\quad
K_{x}=\f V{2\lambda}+\lambda\cH^{reg}
=
v\f{\cos 2\lambda b}{2\lambda}+\frac{\lambda\pi^2_{\phi}}{24 \pi G v\cos^{2}\lambda b}
\,,
\ee
%\,,\quad
\be
J_{z}=\f V{2\lambda}-\lambda\cH^{reg}
=
\f{v}{2\lambda}-\frac{\lambda\pi^2_{\phi}}{24 \pi G v\cos^{2}\lambda b}\,.
\nn
\ee
These generators differ from their purely gravitational counterparts $k_{x},k_{y},j_{z}$ by their matter terms. They nevertheless still form a closed $\SU(1,1)$ algebra, which allows to integrate the flow generated by the complexifier, but also the evolution flow generated by the Hamiltonian constraint, as Lorentz transformations. 
Finally, note that because we have introduced a new volume $V = v\cos^{2}\lambda b$ in the matter sector, a natural question arises. What is the relevant notion of volume we should investigate in the quantum theory : the volume experienced by the matter $V$ or the bare gravitational volume $v$? Intuitively, the new volume $V$ seems to carry more information since it is the one experienced by the matter. Indeed, we shall see that using $V$ instead of $v$ is indeed more suited in the quantum theory.

\smallskip

To summarize, we have a new proposal for a fully regularized effective Hamiltonian for loop quantum cosmology, which leads a closed CVH formula taking into account both mater and gravitational contribution to the Hamiltonian. Both the complexifier and the inverse volume factor are regularized. The CVH algebra is identified as a $\SU(1,1)$ Lie algebra, whose Casimir is determined by the matter energy density. This group structure allows to integrate exactly the flows generated on the phase space by both the complexifier and the full Hamiltonian constraint.

%%%%%%%%%%%%%
\section{Integrating the Dynamics as a $\SU(1,1)$ flow}
%\subsection{Towards a Wick rotation back to the self-dual theory}
%%%%%%%%%%%%%

We can use the $\SU(1,1)$ group structure generated by the CVH algebra to integrate exactly the dynamics of these cosmological models. Indeed we can exponentiate not only the action of  the complexifier but also the action of the Hamiltonian constraint as Lorentz transformations. This allows to describe the cosmological evolution as a $\SU(1,1)$ flow.

We will proceed using two methods, which are ultimately equivalent:
\begin{itemize}

\item {\it Deparametrizing the dynamics in term of the scalar field}:

Using the scalar field as a clock, we solve the Hamiltonian constraint and show that the evolution of the geometry is exactly generated by the complexifier. Not only this allows to solve and integrate the equations of motion, but it also shows the cosmological evolution as a rescaling of the Barbero-Immirzi parameter and underlines the relation between dynamics and scale transformation for gravity.

\item {\it Integrating the flow of the Hamiltonian constraint}:

We exponentiate directly the action of the Hamiltonian constraint, thus describing the evolution in terms of the coordinate time $t$ defined by the choice of lapse $N=1$. The trajectories $v(t)$ and $\phi(t)$ for the geometry and scalar matter can be obtained as a null $\SU(1,1)$ flow. Finally deparametrizing these trajectories to obtain the evolution $v(\phi)$ of the scale factor in terms of the scalar field will lead back to the same equations as in the previous method.

\end{itemize}

%%%%%%%
\subsection{Deparametrizing the dynamics: the Complexifier as Effective Hamiltonian}
%%%%%%%

Since we have one Hamiltonian constraint on a $2\times 2$-dimensional phase space, it imposes both a constraint and a gauge-invariance which reduces the system to a single physical degree of freedom. In practice, we start with both pairs of canonical variables, $(v,b)$ and $(\phi,\pi_{\phi})$, and we deparametrize the system to extract the gauge-invariant physical content of the model: we solve the Hamiltonian constraint and we compute the trajectory $(v(\phi),b(\phi))$ using the scalar field $\phi$ as a clock.

For standard classical FRW cosmology, we solve the Hamiltonian constraint \eqref{Classical} for the scalar field momentum:
\be
\cH^{0}
=
-\f12vb^{2}
+
\f{\pi_{\phi}^{2}}{24\pi G\,v}
=0
\quad\Longrightarrow\quad
\pi_{\phi}=\pm\sqrt{12\pi G}\, vb\,.
\ee
We get two possible branches. We pick the positive branch for the sake of simplicity, but both branches are admissible. This gives the deparametrized Hamiltonian defining the evolution of the geometry variables $(v,b)$ in terms of the scalar field clock $\phi$. The moot point is that this deparametrized Hamiltonian is exactly given (up to a numerical factor) by Thiemann complexifier:
\be
\pi_{\phi}=\sqrt{12\pi G}\, vb
\,=\,
\sqrt{12\pi G}\, C\,.
\ee
First, this means that the effective evolution of the cosmological volume in terms of the scalar matter is given by  canonical transformations rescaling the Barbero-Immirzi parameter. This reflects the fact that the cosmological evolution amounts to scale transformations. Second, it implies that we can integrate the deparametrized dynamics as the flow of the complexifier $C$, which we have already computed in \eqref{Cflow} and induces a simple rescaling of the volume and its conjugate momentum:
\be
\label{vphi-FRW}
v(\phi)
=e^{\phi\sqrt{12\pi G}\{C,\cdot\}} v
=e^{\phi\sqrt{12\pi G}}\,v_{\phi=0}
\,\qquad
b(\phi)
=e^{-\phi\sqrt{12\pi G}}\,b_{\phi=0}\,,
\ee
describing an expanding phase for the volume $v$, with the product $C=vb$ kept constant along each trajectory.
At $\phi\rightarrow-\infty$, the volume starts at 0. It then exponentially increases.
The negative branch, defined by $\pi_{\phi}=-\sqrt{12\pi G}\, C$, would describe the opposite contraction phase.

\smallskip

\begin{figure}[h]
%\begin{figure}[h!]
\includegraphics[height=40mm]{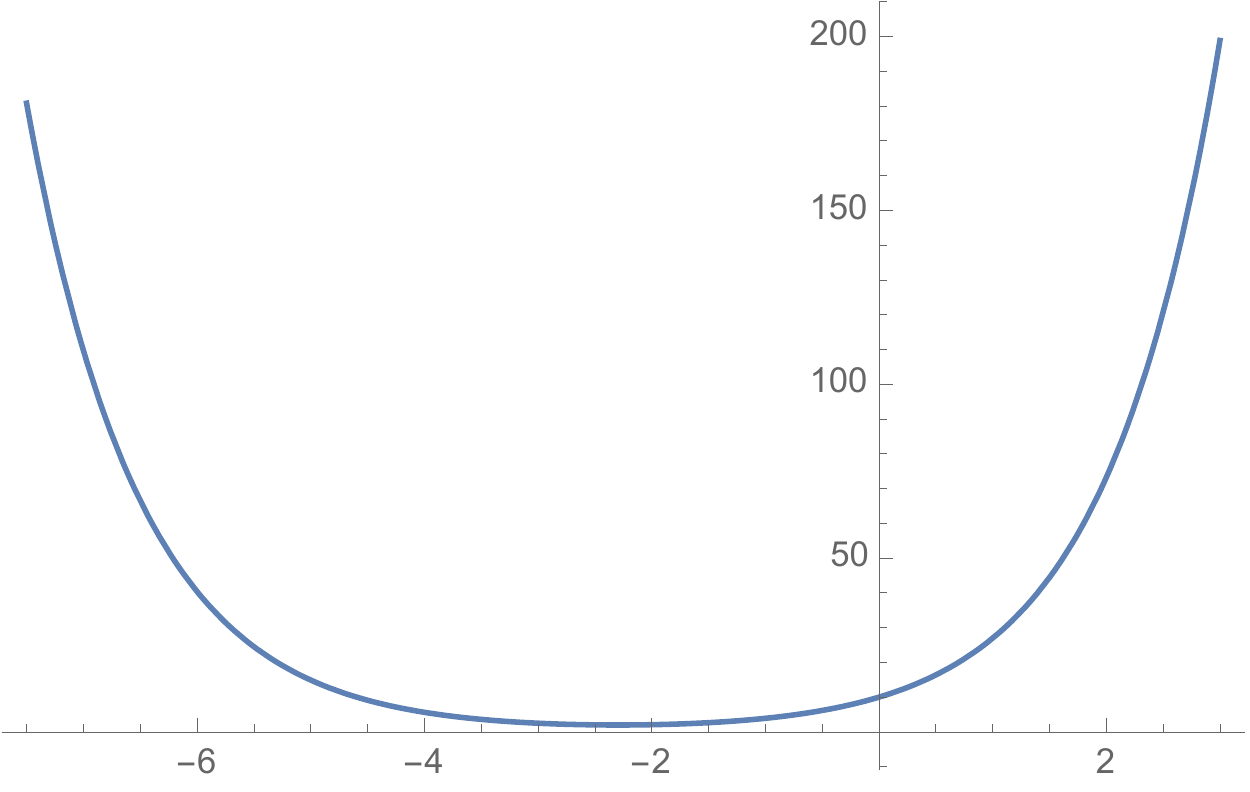}
\hspace*{10mm}
\includegraphics[height=40mm]{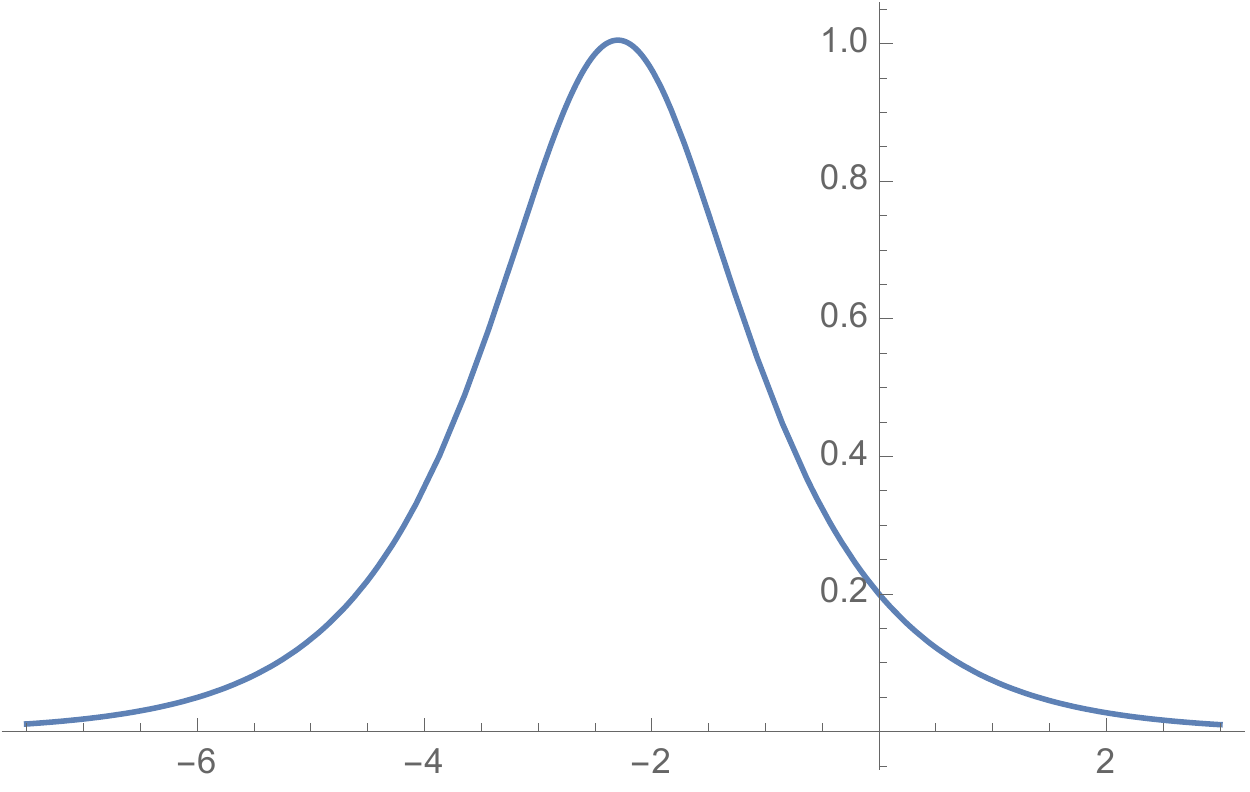}
\caption{\label{vphiplot}Deparametrized dynamics of the volume $v$ and its conjugate angle $\sin2\lambda b$ in terms of the scalar field clock $\phi\sqrt{12\pi G}$ following a hyperbolic evolution, for regularization scale $\lambda=1$ and initial volume $v_{0}=10$ and constant complexifier $\cC=1$. The volume reaches a minimum around $\phi\sqrt{12\pi G}\simeq -2.3$, while $b$ reaches its maximal value. This is the LQC big bounce.}
\end{figure}

What is of special interest is that all these features straightforwardly extend to the effective regularized cosmology of LQC. Indeed, considering the fully regularized Hamiltonian constraint \eqref{Hreg}, with both regularized curvature and regularized inverse volume factor, we find that once again the deparametrized Hamiltonian is the regularized complexifier:
\be
\cH^{reg}
=
\frac{\pi^2_{\phi}}{24 \pi G v\cos^{2}\lambda b} - \frac{1}{2} v \frac{\sin^2{(\lambda b)}}{\lambda^2}
%=
%\f1{\cos^2{(\lambda b)}}
%\,\Bigg{[}
%\frac{\pi^2_{\phi}}{24 \pi G v} - \frac{1}{2} v \frac{\sin^2{(2\lambda b)}}{(2\lambda)^2}
%\Bigg{]}
=0
\quad\Longrightarrow\quad
\pi_{\phi}
=
\pm\sqrt{12\pi G}\,v \frac{\sin{(2\lambda b)}}{2\lambda}
=
\pm\sqrt{12\pi G}\,\cC\,.
\ee
We have two branches as in the standard case. Analyzing the positive branch, the deparametrized evolution is again generated as the flow of the complexifier. This means that we can integrate the deparametrized dynamics as $\SU(1,1)$ boosts. This method works even if we forget the inverse volume regularization and was initially developed in \cite{LivGQCosmo}. We have already computed the flow of the regularized complexifier on the phase space, in equations \eqref{Cregflow} and \eqref{TransfoVH}:
\be
\label{vdeparam}
v(\phi)
%=
%e^{\eta}v +4\lambda^{2}\cH_{g}\sinh\eta
=
e^{\phi\sqrt{12\pi G}}\,v_{0}
-
\f{2\lambda^{2}\cC^{2}}{v_{0}}\sinh\phi\sqrt{12\pi G}
=
\left[v_{0}-\f{\lambda^{2}\cC^{2}}{v_{0}}\right]\,e^{\phi\sqrt{12\pi G}}
+
\f{\lambda^{2}\cC^{2}}{v_{0}}e^{-\phi\sqrt{12\pi G}}\,,
\ee
where $\cC$ is a constant of motion. Since $\cC$ is constant along the trajectory, we can deduce the evolution of the conjugate angle $b$ from the evolution law of the volume:
\be
\cC=v\f{\sin 2\lambda b}{2\lambda}
\quad\Rightarrow\quad
%\sin 2\lambda b(\phi)
%\,=\,
%\f{2\lambda \cC}{v(\phi)}
b(\phi)
\,=\,
(2\lambda)^{-1}\,\arcsin\f{2\lambda \cC}{v(\phi)}
\,.
\ee
An important feature of loop quantum cosmology is that the positive branch is already a superposition of expanding and contracting phases: as illustrated on fig.\ref{vphiplot}, as the scalar field $\phi$ grows from $-\infty$ to $+\infty$, the volume decreases from $\infty$, reaches a minimal value  and then increases back towards $\infty$. The negative branch would give the same formulas.

That the deparametrized dynamics is simply generated by the (regularized) complexifier seems to be a key insight in the theory. Mathematically, it implies that the evolution is exactly integrated in straightforward $\SU(1,1)$ boosts. Physically, it means that the (relational) cosmological dynamic is somehow equivalent to a rescaling of the BI parameter $\gamma$,  which amounts to a scale transformations of the geometry. It is tantalizing to investigate how this generalizes to the framework of full loop quantum gravity. We shall come back on this point in the discussion in section \ref{discuss}.

%%%%%%%
\subsection{Hamiltonian evolution as null $\SU(1,1)$ transformations}
%\subsection{Hamiltonian evolution as $\SU(1,1)$ transformations}
%%%%%%%

Here we develop the second method to integrate the motion of the cosmological model, by directly exponentiating the Hamiltonian constraint. This corresponds to the evolution for the choice of lapse $N=1$. It turns out that the Hamiltonian constraint is a null generator of the CVH $\su(1,1)$-algebra, so the trajectory are generated by null Lorentz transformations.

Let us start with the standard FRW cosmology case. The Hamiltonian constraint is:
\be
\cH^{0}=-\f12vb^{2}+\f{\pi_{\phi}^{2}}{24\pi G\,v}=0\,.
\ee
We compute the Hamilton equations of motion generated by $\cH$:
\be
\pp_{t} \pi_{\phi}=0
\,,\qquad
\pp_{t}\phi=\{\phi,\cH^{0}\}=\f{\pi_{\phi}}{12\pi G\,v}\,,
\ee
\be
\pp_{t}v=\{v,\cH^{0}\}=vb
\,,\qquad
\pp_{t}(vb)=\{vb,\cH^{0}\}=-\cH^{0}\,.
\ee
Since the Hamiltonian constraint vanishes, the dilatation $C=vb$ is constant along a trajectory and the volume $v$ grows linearly with the time $t$:
\be
v(t)= C t +v_{0}\,,
\qquad
\label{depclass}
\cH^{0}=0\Rightarrow C=\pm\f{\pi_{\phi}}{\sqrt{12\pi G}}
\ee
 We  also integrate the matter sector:
\be
\phi(t)= \pm \frac{C}{\sqrt{12\pi G}}\,\ln \left(\f{Ct+v_{0}}{v_{0}}\right)+\phi_{0}
\,.
\ee
We can write the equation of motion for the volume as a Friedman equation:
\be
\f{\pp_{t}v}v
=\f{C}v
\quad\Rightarrow\quad
\left(\f{\pp_{t}v}v
\right)^2
=\ka \rho
\qquad\textrm{with}\quad
\ka=\f{8\pi G}3\,,\quad
\ka\rho
%=\f{C^2}{ v^2}
=\f{\pi_{\phi}^2}{12\pi G v^2}\,,
\ee
where $\rho$ is the energy density of the scalar field.
Combining the two evolution laws $v(t)$ and $\phi(t)$, we get the deparametrized trajectory, describing the evolution of the volume in terms of the scalar field time:
\be
v=v_{0}e^{\pm\sqrt{12\pi G}\,(\phi-\phi_{0})}
\,,\qquad
\f{\rd v}{\rd \phi}
=
\f{\pp_{t}v}{\pp_{t}\phi}
=
\pm\sqrt{12\pi G}\,v
\,,
\ee
with two possible exponential branches. This fits perfectly, as expected, with eqn.\eqref{vphi-FRW} and our previous analysis of the deparametrized dynamics.

\smallskip

We can recover all this from the $\su(1,1)$ algebra structure and derive the Hamiltonian evolution as a  $\SU(1,1)$ flow. The $\su(1,1)$ generators are given in terms of the volume and Hamiltonian constraint:
\be
v=\f{K_{x}+J_{z}}2
\,,\quad
\cH^{0}=K_{x}-J_{z}
\,,\qquad
K_{y}=vb
\,,\quad
K_{x}
=\f v4 +\f{\cH^{0}}2 
%=\f14\bigg{[}
%v+\beta v^{-1}-vb^2
%\bigg{]}
\,,\quad
J_{z}
=\f v4 -\f{\cH^{0}}2 
%=\f14\bigg{[}
%v-\beta v^{-1}+vb^2
%\bigg{]}
\,.
\ee
The Hamiltonian constraint is thus a null-like element in the $\su(1,1)$ Lie algebra. Its exponentiated action $\exp[- t\{\cH,\cdot\}]$ easily translates into a null $\SU(1,1)$ transformation:
\be
\cG_{t}=e^{-i\f t2(\tau_{x}-\tau_{z})}=\id+\f t2(\sigma_{x}+i\sigma_{z})\,\in\SU(1,1)
\,.
\ee
We compute the  action of this $\SU(1,1)$ transformation on $\su(1,1)$ generators, which leads to the evolution of the volume $v$ and dilatation $K_{y}$:
%{\bf (First equation to be removed in final version)}
%\be
%J_{z}(t)=(1+\f{t^{2}}2)J_{z}(0)+t K_{y}(0) -\f{t^{2}}2 K_{x}(0)
%\,,\quad
%K_{+}(t)=K_{+}(0)-itK_{+}(0)+itJ_{z}(0)-\f{t^{2}}2 K_{x}(0)+\f{t^{2}}2 J_{z}(0)\,,
%\nn
%\ee
\be
v(t)=v(0)+tK_{y}(0)-t^{2}\cH^{0}(0)
\,,\qquad
K_{y}(t)=K_{y}(0)-t\cH^{0}(0)\,,
\ee
where we must remember to keep the Hamiltonian constraint $\cH^{0}(0)=0$ as initial conditions for all trajectories. So the volume grows linearly with the time coordinate $t$ and the complexifier $C=K_{y}=vb$ remains constant along the trajectories. And the null $\SU(1,1)$-flow fits exactly with the trajectories integrated by hand \eqref{depclass}.

\smallskip

The efficiency of the $\SU(1,1)$ framework is that the same group element $\cG_{t}$ can now be used to integrate the equations of motion for the effective Hamiltonian of loop quantum cosmology. Indeed we have the exact same $\SU(1,1)$ structure with the same expression of the Hamiltonian constraint as a null-like $\su(1,1)$ generator.

For the usual effective Hamiltonian $\cH$, given in \eqref{effH}, this technique only allows to exponentiate the action of the gravitational Hamiltonian. Indeed the inverse volume factor $v^{-1}$ appearing in the matter Hamiltonian does not belong to the $\su(1,1)$ algebra. It affects the dynamics of the gravitational sector and precludes the integration of the Hamiltonian as a $\SU(1,1)$ flow on the phase space.

However, this feature is entirely corrected in our new proposal for an effective  Hamiltonian $\cH^{reg}$, given in \eqref{Hreg}, which takes into account the regularization of the inverse volume factor. The exponentiated action $\exp[- \tau\{\cH,\cdot\}]=\exp[- t\{K_{x}-J_{z},\cdot\}]$, with the normalized time $(2\lambda)\,t=\tau$, is computed as above. Imposing that the Hamiltonian constraint is satisfied $\cH^{reg}=0$, we get the following trajectory:
\be
\cC, \pi_{\phi} \,\,\textrm{constant}
\,,\quad
V(t)=V_{0}+t\cC
\,,\quad
\pp_{t}\phi
=
\f{\pi_{\phi}}{12\pi G\,V}
\,.
\ee
The values of the complexifier $\cC$ and the matter energy density are related to each other by the Hamiltonian constraint:
\be
\label{effeH}
\cH^{reg}=0
\,\Rightarrow\quad
\pi_{\phi}^{2}=12\pi G \,\cC^{2}\,,
\ee
exactly as the standard FRW case. Again, the cosmological dynamic is given by the complexifier, generating the rescaling of the BI parameter. This suggests a nice interplay between the cosmological dynamic and a renormalization process of $\gamma$. This will be discussed in the last section. From there, we can compute the volume:
\be
v=V+\f{\lambda^{2}\cC^{2}}V
\,,\qquad
V=\f{v+\sqrt{v^{2}-4\lambda^{2}\cC^{2}}}2\,.
\ee
One can integrate for the matter field $\phi$ in terms of the time $t$ and express this directly in terms of the regularized volume $V$:
\be
\phi= \pm \frac{C}{\sqrt{12\pi G}}\ln\left( \f V {V_{0}}\right) +\phi_{0}
\,,\qquad
V=V_{0}e^{\pm \sqrt{12 \pi G}\,(\phi-\phi_{0})}\,,
\ee
which allows to get the (deparametrized) trajectory of the volume $v$ in terms of the scalar field time $\phi$:
\be
v
=
V_{0}e^{\pm \sqrt{12 \pi G}\,(\phi-\phi_{0})}
+
\f {\lambda^{2}\pi_{\phi}^{2}}{12\pi G V_{0}}e^{\mp \sqrt{12 \pi G}\,(\phi-\phi_{0})}\,,
\ee
with a mixing of the two exponential branches. And we recover the results \eqref{vdeparam} which we got earlier by integrating directly the deparametrized Hamiltonian:
at the end of the day,  the null flow generated by the Hamiltonian constraint fits with the boost flow of the (regularized) complexifier describing the deparametrized dynamics.

We can go further and compute the modified Friedman equations for the regularized cosmological model:
\be
v=V+\f{\lambda^{2}\cC^{2}}V
\quad\Rightarrow\quad
\f{\pp_{t}v}{v}
=\f{\pp_{t}V}{V}  \left(\f{V^{2}-\lambda^{2}\cC^{2}}{V^{2} +\lambda^{2}\cC^{2}}\right)\,.
\ee

The key point is that this derivative can vanish, when $V^2=\lambda^2\cC^2$, or equivalently when $v=2\lambda\cC$. This gives a critical density:
\be
\ka\rho_{c}=\f1{4\lambda^2}\,,
\ee
defined simply in terms of the regularization scale. This typically signals a bounce occurring at a minimal (dynamical) volume, which depends on the matter momentum, as expected in loop quantum cosmology:
\be
v_{bounce}=2\lambda\f{\pi_{\phi}}{\sqrt{12 \pi G}}\,.
\ee
More generally, we can write the exact modified Friedman equation:
%\be
%\f{\pp_{t}v}{v}
%=\f\cC v\,\left(1-\f{\lambda^2\cC^2}{V^2}\right)
%=\f\cC v\,\left(1-\left(\f vV\right)^2\f{\lambda^2\cC^2}{v^2}\right)
%\quad\textrm{with}\quad
%\f vV
%=
%\f2{1+\sqrt{1-\f{4\lambda^2\cC^2}{v^2}}}
%\ee
%\be
%\left(\f{\pp_{t}v}{v}\right)^2
%=\ka\rho\left(1-\left(\f vV\right)^2\f{\rho}{4\rho_{c}}\right)
%\quad\textrm{with}\quad
%\f vV
%=
%\f2{1+\sqrt{1-\f{\rho}{\rho_{c}}}}
%\ee
\be
\f{\pp_{t}v}{v}
=\f\cC v\,\left(1-\f{\lambda^2\cC^2}{V^2}\right)
\quad\Rightarrow\quad
\left(\f{\pp_{t}v}{v}\right)^2
=\ka\rho\left[1-\f1{\left(1+\sqrt{1-\f{\rho}{\rho_{c}}}\right)^2}\f{\rho}{\rho_{c}}\right]^2\,,
\ee
where we get a slight correction to the usual LQC Friedman equation. Of course, one should keep in mind that we could alway choose a slightly different lapse, and thus time parameter, to compensate for this deviation. But we insist that this is an exact result coming straight fro the flow of the Hamiltonian constraint of our regularized effective Hamiltonian (with both regularized volume and complexifier). Note that at leading order in $\frac{\rho}{\rho_c} \ll 1$, one the following modified Friedman equation
\be
 \left(\f{\pp_{t}v}{v}\right)^2 = \kappa \rho \left[ 1 - \frac{\rho}{2\rho_c} - \frac{3}{16} \frac{ \rho^2}{\rho^2_c} + o\big{(} \frac{\rho^3}{\rho^3_c}\big{)} \right]\,
\ee

This concludes the classical analysis and we are now ready to proceed to the quantization of the theory.

%%%%%%%%%%%%%%
\section{Quantizing the $\SU(1,1)$ CVH Algebra: Back to Loop Quantum Cosmology}
%\section{Back to Loop Quantum Cosmology: Quantizing the $\SU(1,1)$ CVH Algebra}
%%%%%%%%%%%%%%

We now proceed to the quantization of the cosmological phase space, focusing on the $\SU(1,1)$ group structure provided by the CVH algebra. So we define the Hilbert space as the theory as a $\SU(1,1)$ representation. Following the structures of the classical theory, the volume, complexifier and Hamiltonian constraint will be quantized as the $\su(1,1)$-generators  and the $\SU(1,1)$ representation will be chosen by the value of the Casimir given in terms of the volume gap $v_{m}$ and matter energy density $\pi_{\phi}$.

We will distinguish two ways to proceed. On the one hand, we can focus on the gravitational CVH algebra, thus quantize the gravitational sector separately from the matter sector: the $\SU(1,1)$ representation will represent only the gravitational degrees of freedom. This is the usual route taken by loop quantum cosmology. On the other hand, we can take a more global point of view exploiting that the extension of the CVH algebra taking into account the matter contribution to the Hamiltonian constraint. This will encode both gravitational degrees of freedom and matter degrees of freedom mixed together right from the start in the $\SU(1,1)$ group structure and generators.

We start by giving the Casimir equation and identify the irreducible representations of the $\SU(1,1)$ group that we must use. Then we use coherent states \`a la Perelomov to describe the cosmological quantum  states, in both schemes. The complexifier will always become a Hermitian operator $\hat{\cC}$ generating unitary transformations on the Hilbert space.

%%%%%
\subsection{Selecting the $\SU(1,1)$ representation: the Casimir equation}
%%%%%

Unitary irreducible representation of the $\su(1,1)$ Lie algebra are easy to describe in the standard basis diagonalizing both the Casimir operator and the compact rotation generator $\hat{\cJ}_{z}$. We introduce the basis states $|\mathfrak{C},m\ra$ where $\mathfrak{C}$ is the value of the Casimir operator $\hat{\mathfrak{C}}$ and the integer\footnotemark $m\in\Z$ gives the discrete eigenvalues of $\hat{J}_{z}$. Then the $\su(1,1)$-generators are well-known to act as:
\footnotetext{
In fact, we have two cases, either $m\in\Z$ or $m\in\Z+\f12$. Here we only consider the integer case $m\in\Z$ for the sake of simplicity, which actually corresponds to unitary representations of $\SO(2,1)$. Working with the half-integer representations would not change anything to our quantization scheme.
}
\be
\begin{array}{lcl}
\hat{\mathfrak{C}}\,|\mathfrak{C},m\ra
&=&
\Big{[}\hat{\cJ}_{z}^2-\f12\hat{\cK}_{+}\hat{\cK}_{-}-\f12\hat{\cK}_{-}\hat{\cK}_{+}\Big{]}
\,|\mathfrak{C},m\ra 
=
\mathfrak{C}\,|\mathfrak{C},m\ra \\
\hat{\cJ}_{z}\,|\mathfrak{C},m\ra
&=&
m\,|\mathfrak{C},m\ra  \\
\hat{\cK}_{+}\,|\mathfrak{C},m\ra
&=&
\sqrt{m(m+1)-\mathfrak{C}}\,|\mathfrak{C},m+1\ra \\
\hat{\cK}_{-}\,|\mathfrak{C},m\ra
&=&
\sqrt{m(m-1)-\mathfrak{C}}\,|\mathfrak{C},m-1\ra
\end{array}
\ee
We distinguish two series of unitary representations. The space-like representations have a negative  Casimir $\mathfrak{C}< 0$ and the discrete label $m$ runs through the whole set of integers $\Z$. We usually write $\mathfrak{C}=-s^2$ and these form the continuous series\footnotemark{} of representations $\cC_s$.
\footnotetext{
Actually we usually distinguish two series of representations with negative Casimir. The principal continuous series with $\mathfrak{C}=-s^2-\f14< -\f14$ come in the Plancherel decomposition for square-integrable functions on $\SU(1,1)$, while the representations with $-\f14<\mathfrak{C}<0$ are called the exceptional continuous series.
Moreover, there actually exists two continuous series of representations $\cC^{\epsilon}_s$ labeled by the parity $\eps=\pm$, the positive parity $\eps=+$ corresponding to $m\in\Z$  and $\eps=-$ corresponding to $m\in\Z+\f12$.
}
The time-like representations have a positive or vanishing Casimir. The Casimir can only take discrete values, $\mathfrak{C}=j(j-1)\ge0$ for $j\ge 1$. The weights $m$ are not allowed to run through the whole $\Z$. We distinguish the positive discrete series $\cD^{j}_{+}$ whose Hilbert space is spanned by the basis states with $m\ge j$ and the negative discrete series $\cD^{j}_{-}$ whose Hilbert space is spanned by the basis states with $m\le -j$.

\smallskip

Now the value of the Casimir and thus the choice of representation will depend on the details of the cosmological model. Let us start with the case of standard FRW cosmology. The CVH algebra for the gravitational sector already forms a $\su(1,1)$ algebra whose generators are:
\be
j_{z}= v\left[1+\f{b^2}4\right]
= v- \f12\cH^0_{g}
\,,\qquad
k_{x}=v+\f12\cH^0_{g}
\,,\qquad
k_{y}=vb=C
\,,
\nn
\ee
and the Casimir vanishes,
\be
\mathfrak{C}
=j_{z}^2-k_{x}^2-k_{y}^2
= -  2v\cH^{0}_{g} - C^{2}  =0
\,.
\nn
\ee
We would like to use a null-like representation, which can correspond either to $\mathfrak{C}\rightarrow 0$ limit of the series of continuous representations or to the $j=1$ case of the discrete series, which has a vanishing Casimir. Considering that the generator $j_{z}$ is always positive at the classical level, it seems natural to opt for the latter and quantize the theory using the representation $\cD^{j=1}_{+}$  of the discrete positive series, for which the eigenvalues of $\hat{j}_{z}$ are always positive $m\ge 1$. Let us however point out that here $\hat{j}_{z}$ is not the volume operator.

Now if we consider the CVH algebra extended to take into account both the gravitational and the matter sector, the generator of the $\su(1,1)$ algebra acquire a matter term, but retain a similar definition as previously:
\be
J_{z}
= v- \f12\cH^0
= v\left[1+\f{b^2}4\right]-\f{\pi_{\phi}^2}{48\pi G v}
\,,\qquad
K_{x}=v+\f12\cH^0
\,,\qquad
K_{y}=C
\,,
\nn
\ee
with a negative Casimir:
\be
\mathfrak{C}
=J_{z}^2-K_{x}^2-K_{y}^2
= -  2v\cH^{0} - C^{2} 
=-\f{\pi_{\phi}^2}{12\pi G v}<0
\,.
\nn
\ee
This means that the whole system gravity plus matter field should be quantized as a space-like representation from the continuous series.

Let us now turn to the regularized Hamiltonians for loop quantum cosmology.

\begin{itemize}

\item {\it CVH algebra for the gravitation sector:}

The CVH algebra for the regularized Hamiltonian is again written as a $\su(1,1)$-algebra with generators:
\be
j_{z}=\f1{2\lambda}v
\,,\qquad
k_{\pm}=v\,e^{\pm 2i\lambda}
%k_{\pm}=\sqrt{v^2-v^2_{m}}\,e^{\pm 2i\lambda}\,,
%\nn
%\ee
%\be
\,,\qquad
\cC=k_{y}=\f1{2i}(k_{+}-k_{-})
\,,\qquad
\cH_{g}=\f1{2\lambda}(k_{x}-j_{z})\,,
\ee
where $\lambda$ is the regularization scale. The Casimir still vanishes:
\be
\mathfrak{C}
=j_{z}^2-k_{x}^2-k_{y}^2
= -  2V\cH_{g} - C^{2} 
= 0
\,,
\qquad\textrm{with}\quad
V=\f12(j_{z}+k_{x})
\,,
\ee
where $V$ is the regularized volume.
Since the volume $v$ should remain positive, we must choose representations from the discrete positive series $\cD^{j}_{+}$. The Casimir gets then naturally regularized to a positive value, $\mathfrak{C}=j(j-1)\ge0$, where we can still get a vanishing Casimir at the quantum level for the minimal value of the spin $j=1$. 

Here the volume operator $\hat{v}=2\lambda \, \hat{j}_{z}$ has a discrete positive spectrum  $2\lambda m$ with $m\ge j$. This leads back to the usual framework for loop quantum cosmology, which quantizes separately the gravitational and matter sectors.
We will describe in details the action of the complexifier and the definition of coherent states in the next section.

So the spin $j\ge 1$ encodes the minimal value possible of the volume, which can never reach 0. It is actually possible to introduce such a minimal volume $v_{m}$ directly in the classical phase space, reflecting more closely the features of the quantum theory, as explained in appendix \ref{vmin}.

\item {\it CVH algebra for the full  system, gravity coupled to matter field:}

We have shown earlier that the CVH algebra can be extended to the whole system by properly regularizing the inverse volume factor of the matter term of the Hamiltonian constraint, simply replacing the volume $v$ by the regularized volume $V$. This allows for a homogeneous behavior of the whole Hamiltonian constraint under the action of the (regularized) complexifier. The $\su(1,1)$ generators are defined as for the pure gravitation sector:
\be
K_{y}=\cC
\,,\qquad
\cH=\f1{2\lambda}(K_{x}-J_{z})
\,,\qquad
V={\lambda}(K_{x}+J_{z})\,,
\ee
 with a Casimir given by the matter energy density,
\be
\label{FullCaz}
\mathfrak{C}_{\su_{1,1}} =J_{z}^{2} - K_{x}^{2}- K_{y}^{2} = - \f{\pi_{\phi}^{2}}{12\pi G} 
%\mathfrak{C}_{\su_{1,1}} =J_{z}^{2} - K_{x}^{2}- K_{y}^{2} = v_{m}^{2} - \f{\pi_{\phi}^{2}}{12\pi G} 
\,.
\ee 
Thus the matter energy produces a negative Casimir and leads to a quantization using a $\SU(1,1)$-representation from the continuous series.
Once the $\SU(1,1)$-representation is selected, $J_{z},K_{x},K_{y}$  become quantum operators acting on that representation. It is important to keep in mind that $J_{z}$ is not the volume $v$ anymore, but a more complicated function involving $v$, $b$ and the matter momentum $\pi_{\phi}$. So at the quantum level, it will not be straightforward anymore to identity a volume operator $\hat{v}$. The discrete spectrum of $\hJ_{z}$ now corresponds to the quantization of a more complicated observable mixing the geometry and the matter.

At this point, we have two important remarks.
First, this quantization scheme involves the gravity and matter sectors at the same  level, mixing them. From this perspective, a legitimate question is whether this $\SU(1,1)$-quantization scheme will lead back to the same quantum theory as in standard LQC. For instance, we don't get a straightforward discrete spectrum for the volume as when quantizing separately the gravity and matter sectors.
Of course, the volume is not a physical observable (it does not commute with the Hamiltonian constraint) and it is not clear that the discrete spectrum of a kinematical operator stays relevant at the physical level. In fact, LQC does mix the gravity and matter sectors when solving the Hamiltonian constraints for physical states and, at the end of the day, it is likely that the Hilbert space of physical states for LQC and for our $\SU(1,1)$-quantization will be the same, although this explicitly remains to be checked. We postpone this to future investigation and this will be the necessary next step of this program.

Second, the matter momentum $\pi_{\phi}$ is a constant of motion, since it commutes with the Hamiltonian constraint. So we can fix it and study the dynamics of the cosmological model at fixed $\pi_{\phi}$, therefore working with a single irreducible representation of $\SU(1,1)$. Nevertheless, if we would like to define a Hilbert space for arbitrary quantum states for both the geometry and the matter field, then we need to consider a direct sum over all those $\SU(1,1)$ representations allowing for arbitrary wave-functions of the matter scalar field.

\end{itemize}

%%%%%
\subsection{The gravitational sector: coherent states and unitary Immirzi transformations}
%%%%%

In this section, we focus on the gravitational sector, defining the $\SU(1,1)$-representation encoding the CVH algebra at the quantum level, introducing the corresponding coherent states of geometry and deriving the action of the complexifier on that Hilbert space. This can be considered as the quantization of the gravitational sector, but also as the quantization of the deparametrized theory. From that latter perspective, the complexifier actually defines the dynamics of the geometry in terms of the scalar field clock.

As we have described above, we consider a $\SU(1,1)$-representation from the positive discrete series $\cD^j_{+}$ for an integer $j\ge 1$. The Hilbert space is spanned by basis states $|j,m\ra$ with $m\in\N$, $m\ge j$. The $\su(1,1)$-generators allow to define the quantum operators corresponding to the volume, the complexifier and the gravitational Hamiltonian:
\be
\hat{v}=2\lambda\hj_{z}
\,,\quad
\hat{\cC}=\hk_{y}
\,,\quad
\hat{\cH}_{g}=\f1{2\lambda}(\hk_{x}-\hj_{z})\,.
\ee
We also define the regularized volume operator $\hV=\lambda(\hj_{z}+\hk_{x})$.
These are all Hermitian operators since we are working with a unitary representation of $\SU(1,1)$. In particular, both the complexifier $\hat{\cC}$ and the gravitational Hamiltonian $\hat{\cH}_{g}$ generate unitary transformations. Moreover the volume $\hat{v}$ has a discrete spectrum $2\lambda m$ with a minimal non-zero volume $2\lambda j$. The Casimir operator is constant on each representation and its value depends only on the value of $j$:
\be
\label{casimir-q}
\hat{\mathfrak{C}}=\hj_{z}^2-\hk_{x}^2-\hk_{y}^2
=-\,\Big{[}\hat{V}\hat{\cH}_{g}+\hat{\cH}_{g}\hat{V}+\hat{\cC}^2\Big{]}
=j(j-1)\,.
\ee

It is possible to define coherent states \`a la Perelomov \cite{LivGQCosmo}. These will be the coherent states for the geometry in LQC:
\begin{align}
|j, z \ra   =  \sum^{+ \infty}_{m \in j + \mathbb{N}} \; \sqrt{\frac{(m+ j + 1)! }{(m - j)! (2j + 1)!}} \frac{(z^1)^{m-j}}{(\bar{z}^0)^{m + j +2}}  \; | j, \; m \ra 
\end{align}
Those coherent states are labelled by a spinor $z \in \mathbb{C}^2$, whose components $z^0$ and $z^1$ are arbitrary complex numbers. The states are well-defined as long as $|z^0| > | z^1|$ and their norm is:
\be
\la j,z | j,z \ra 
%= \frac{1}{( |z^0|^2-  |z^1|^2)^{N+ 1}}
=  \frac{1}{(2 L)^{2j}} 
\,,\qquad
L=\f12\,\big{(}|z^0|^2-|z^1|^2\big{)}
\ee
The key property of those states is that they are covariant under $\SU(1,1)$ transformations, meaning that a finite $\SU(1,1)$ group element acts on the quantum state $|j, z \ra$ by acting directly on its spinor label as a 2$\times$2 matrix:
 \be
 U  | j ,z \ra =  | j,U \triangleright z \ra\,.
 %=  | j \; z_U \ra
 \ee
Moreover these coherent states are semi-classical states with minimal spread  \cite{LivGQCosmo} and we know the expectation values of the $\su(1,1)$ generators\footnotemark{}.

\footnotetext{
We can compute the expectation values of the $\su(1,1)$ generators by straightforward calculations:
\be
\f{\la j,z | J_{z}|j,z \ra }{\la j,z | j,z \ra}
%=   j\,\frac{|z^0|^2+|z^1|^2}{|z^0|^2-  |z^1|^2}
=  j\,\frac{v_{z}}{L} 
\,,\qquad
v_{z}=\f12\,\big{(}|z^0|^2+|z^1|^2\big{)}\nn
\ee
\be
\f{\la j,z |K_{\pm}|j,z \ra }{\la j,z | j,z \ra}
=  j\,\frac{v_{\pm}}{L} 
\,,\qquad
v_{+}=\bz^0\bz^1
\,,\quad
v_{-}=z^0 z^1=\bar{v}_{+}\nn
\ee
with $v_{z}^2-v_{+}v_{-}=L^2>0$
}
This allows to compute the expectation value of the volume operator,
\be
\la\hat{v}\ra
=\f{\la j,z | \hat{v}|j,z \ra }{\la j,z | j,z \ra}
=2\lambda\,\la \hj_{z}\ra
=2\lambda j\,\frac{|z^0|^2+|z^1|^2}{|z^0|^2-  |z^1|^2}
\ee
as well as of the regularized volume operator, complexifier and gravitational Hamiltonian:
\be
\label{Vmean}
\la \hat{V}\ra
=\lambda\,\la \hk_{x}+\hj_{z}\ra
=\lambda\,j\,
\frac{z^0 z^1+\bz^0\bz^1+|z^0|^2+|z^1|^2}{|z^0|^2-  |z^1|^2}
\ee
\be
\label{Cmean}
\la\hat{\cC}\ra
=\la \hk_{y}\ra
=\f1{2i}\la \hk_{+}-\hk_{-}\ra
=-ij\,\frac{\bz^0\bz^1-z^0 z^1}{|z^0|^2-  |z^1|^2}
\ee
\be
\label{Hmean}
\la \hat{\cH}_{g}\ra
=(2\lambda)^{-1}\,\la \hk_{x}-\hj_{z}\ra
=\f{j}{2\lambda}\,
\frac{z^0 z^1+\bz^0\bz^1-|z^0|^2-|z^1|^2}{|z^0|^2-  |z^1|^2}
\ee
These expectation values also satisfy a semi-classical Casimir equation, which almost matches the exact quantum Casimir formula \eqref{casimir-q} up to a sub-leading term:
\be
\label{casimir-cl}
-\Big{[}2\la V \ra\la \hat{\cH}_{g}\ra+\la\hat{\cC}\ra^2\Big{]}
=j^2>0
\,.
%-( 2\la V \ra\la \hat{\cH}_{g}\ra+\la\hat{\cC}\ra^2)=j^2\,,
\ee

\smallskip

Using the transformation law of the coherent states under the $\SU(1,1)$-action, we can exponentiate exactly the action of the complexifier $\hat{\cC}$. This gives unitary transformations, implementing shifts in the BI parameter at the quantum level. Immirzi transformations for a real parameter $\eta\in\R$ are given by pure boosts, as in the classical theory:
\be
\hat{W}_{\eta} | j,z \ra
 = e^{i\eta \hat{\cC}}| j,z \ra
 = | j, W_{\eta} \triangleright z \ra
 = | j, e^{i\eta \tau_{y}} \triangleright z \ra
 \,,\qquad
 e^{i\eta \tau_{y}}
 = \mat{cc}{\cosh{\f\eta2} &  \sinh{\f\eta2} \\  \sinh{\f\eta2} & \cosh{\f\eta2} }
\,,
\ee
\be
\mat{c}{z_{\eta}^0\\ \bz_{\eta}^1}=
e^{i\eta \tau_{y}} \triangleright z=
\mat{cc}{\cosh{\f\eta2} &  \sinh{\f\eta2} \\  \sinh{\f\eta2} & \cosh{\f\eta2} }
\mat{c}{z^0\\ \bz^1}
\,,
\ee
where one should pay special care to taking the complex conjugate of the second component of the spinor $z$.
This action of the complexifier at the quantum level does not change the representation $j$ and so does not affect the volume gap.  Moreover the flow induced on the spinor $z$ keeps $L=(|z^0|^2-|z^1|^2)$ fixed, as well as $ (\bz^0\bz^1-z^0 z^1)$, which corresponds to the fact that the action of $\hat{W}_{\eta} = e^{i\eta \hat{\cC}}$ as a unitary operator leaves invariant both the norm of the coherent state $\la j,z | j,z \ra $ and its expectation value $\la \hat{\cC}\ra$ given by \eqref{Cmean}.

\smallskip

Not only does this give the flow generated by the complexifier as a $\SU(1,1)$ boost action on the Hilbert space, but this describes the cosmological evolution of the geometry in terms of the scalar field clock in the deparametrized formulation. To make this more explicit, we can actually retrieve the value of the scalar field momentum $\pi_{\phi}$ from the coherent state data $z$.
The matter energy density does not come in the gravitational part of the Hamiltonian constraint, but balances it out in the full Hamiltonian constraint:
\be
\cH^{reg}=\f{\pi_{\phi}^2}{24\pi G V} +\cH_{g}\,.
\ee
So the natural proposal recovers $\pi_{\phi}$ from the expectation value $\la V \ra\la \hat{\cH}_{g} \ra$ given by \eqref{Vmean} and \eqref{Hmean} and set at the quantum level:
\be
\f{\pi_{\phi}^2}{12\pi G}
\equiv\,-2\,\la V \ra\la \hat{\cH}_{g} \ra
=
{j^2}\,
\frac{\big{(}z^0 z^1+\bz^0\bz^1\big{)}^2-\big{(}|z^0|^2+|z^1|^2\big{)}^2}{\big{(}|z^0|^2-  |z^1|^2\big{)}^2}\,.
\ee
Since $-2\,\la V \ra\la \hat{\cH}_{g} \ra=j^2+\la \hat{\cC}\ra^2$ is constant along the orbits generated by action of $ \hat{\cC}$, this is indeed a consistent definition: the scalar field momentum $\pi_{\phi}$ remains constant during the (deparametrized) cosmological evolution as expected.
 
This also clarifies the classical interpretation of the spinor label of the coherent states. Out of the four real components  of $z\in\C^2$, we can remove the norm $L$ which is simply a normalization factor and we are left with three components that determine the semi-classical values of the volume $v$, the complexifier $\cC$, the gravitational Hamiltonian $\cH_{g}$ and the scalar field momentum $\pi_{\phi}$, which are all related by the Hamiltonian constraint $\cH^{(reg)}=0$.
 
 \smallskip
 
 We conclude with two remarks. First, the volume $\hat{v}$ has a discrete spectrum. It does not simply get rescaled under the action of the complexifier:
\begin{align}
\hat{v} \;\;\; \rightarrow \;\;\; \hat{v}_{\eta} = W_{\eta} \hat{v} W^{\dagger}_{\eta}
= e^{\eta} \hat{v} + 4\lambda^2 \sinh{\eta} \; \hat{\cH}_g
\,,
\qquad \qquad
[\hat{v}, \hat{v}_{\eta} ] \ne 0\,.
\end{align}
The initial volume  operators $\hat{v}$  and the Immirzi-shifted volume operator $\hat{v}_{\eta}$ still have the same spectrum, and in particular the same minimal eigenvalue, i.e. the volume gap is not affected by the action of the complefixier. However they do not commute anymore: the exponentiated action of the complexifier does not send a volume eigenstate onto another volume eigenstate.
On the other hand, the regularized volume operator $\hat{V}$ has a continuous spectrum and does get simply rescaled under the action of the complexifier:
\be
\hat{V} \;\;\; \rightarrow \;\;\; \hat{V}_{\eta} = W_{\eta} \hat{V} W^{\dagger}_{\eta}
= e^{\eta} \hat{V}\,.
\ee

A second remark concerns which Hamiltonian constraint to use. Here we have focused on the regularized Hamiltonian $\cH^{reg}$ which uses the regularized inverse volume factor  $V^{-1}$ in the matter term. We have seen above that it this is fully consistent with the action of the complexifier as deparametrized Hamiltonian for the cosmological evolution. We could also follow the usual procedure used in LQC to keep the standard inverse volume factor $v^{-1}$ in the Hamiltonian constraint and define a Hamiltonian constraint operator $\hat{\cH}$ by defining an inverse volume operator $\widehat{v^{-1}}$ at the quantum level.  This is easily done, without any ambiguity, since the volume operator is Hermitian with a strictly positive discrete spectrum. Nevertheless, this would break the elegant relation between the Hamiltonian constraint for coupled gravity and matter and the deparametrized Hamiltonian given by the (regularized) complexifier and we believe this relation to have a deep fundamental meaning.

%%%%%
\subsection{Coupled geometry and matter: Solving the Hamiltonian Constraint}
%%%%%

In this last section, we describe the quantization of the coupled gravity+matter system as a $\SU(1,1)$ representation. The CVH algebra is extended to take into account the whole Hamiltonian constraint with both gravity and matter terms.
%Let us set $v_{m}=0$ for the sake of simplicity.
The main goal here is to investigate the consequences of including the matter term and in particular to understand how to deal with a space-like $\SU(1,1)$-representation from the continuous series.

The $\su(1,1)$generators are given by:
\be
\cC=K_{y}
\,,\quad
V=\lambda(K_{x}+J_{z})
\,,\quad
\cH^{reg}=(2\lambda)^{-1}(K_{x}-J_{z})
\,.
\nn
\ee
Both $J_{z}$ and $K_{x}$ contain a matter term. In particular, $J_{z}$ is not simply the volume $v$ anymore, it depends also on its conjugate angle $b$ and on the scalar field momentum $\pi_{\phi}$:
\be
J_{z}
=\f{V}{2\lambda}-\lambda\cH^{reg}
=
\f{v}{2\lambda}-\f{\lambda\pi_{\phi}^2}{24\pi Gv \cos^2\lambda b}\,.
\ee
The $\su(1,1)$-Casimir is now negative and its numerical value depends on $\pi_{\phi}$:
\be
\mathfrak{C}=-2V\cH^{reg}-\cC^2=-\f{\pi_{\phi}^2}{12\pi G}\,.
\ee

At the quantum level, this selects the $\SU(1,1)$-representation from the continuous series $\cC_{s}$, such that the label $s$ is given by the scalar field momentum, $s^2={\pi_{\phi}^2}/{12\pi G}$. The eigenvalues of $\hJ_{z}$ are unbounded and the Hilbert space is spanned by states $|s,m\ra$ with $m\in\Z$. So we do not get a discrete spectrum for the volume $v$. Actually it is not clear how to define an operator $\hat{v}$, since the generator $J_{z}$ already contains a combination of $v$ and $b$.
The complexifier $\hat{\cC}=\hK_{y}$ is Hermitian and generates unitary transformations on the Hilbert space, without changing the representation label $s$.

In this context, it is possible to define $\SU(1,1)$-coherent states (i.e. that transform covariantly under the $\SU(1,1)$-action) for those space-like representations using a spinorial reformulation, to similarly the construction already done for time-like representations in \cite{LivGQCosmo} and used above for the quantization of the gravity sector. The interested reader can find the relevant spinorial realization of the $\su(1,1)$-algebra  in appendix \ref{newsu11}. This would provide semi-classical states transforming in a straightforward fashion under the exponentiated action of the complexifier $\hat{\cC}$ and of the Hamiltonian constraint $\hat{\cH}^{reg}$.
But here, we are more interested in identifying physical states, satisfying the Hamiltonian constraint $\cH^{reg}=0$. This means finding states $|\phi\ra$, satisfying $(\hK_{x}-\hJ_{z})\,|\phi\ra=0$. Decomposing such a state on the $m$-basis, $|\phi\ra=\sum_{m}\phi_{m}\,|m\ra$, leads to a second order recursion relation on its coefficients:
\be
\hK_{x}\,|\phi\ra=\hJ_{z}\,|\phi\ra
\quad\Longleftrightarrow\quad
\forall m\,,\,\,
2m\phi_{m}=\phi_{m-1}\sqrt{s^2+m(m-1)}+\phi_{m+1}\sqrt{s^2+m(m+1)}
\,,
\ee
or explicitly,
\be
\phi_{-1}+\phi_{1}=0
\,,\quad
2\phi_{1}=\phi_{0}s+\phi_{2}\sqrt{s^2+2}
\,,\quad
4\phi_{2}=\phi_{1}\sqrt{s^2+2}+\phi_{3}\sqrt{s^2+6}
\,,\quad\dots
\ee
So we get a two-dimensional space of physical states satisfying $\hat{\cH}^{reg}\,|\phi\ra=0$, labeled by the initial conditions $\phi_{0}$ and $\phi_{1}$.

We could also look for physical states in the continuous basis of eigenstates of the generator $\hK_{y}$, which diagonalizes the complexifier $\hat{\cC}$ (see e.g. \cite{Davids:2000kz} for details). An important feature of the quantum theory is that $\hat{\cC}$ commutes with the Hamiltonian constraint on physical states:
\be
[\hK_{x}-\hJ_{z},\hK_{y}]=-i(\hK_{x}-\hJ_{z})\,,
\qquad\textrm{which means that}\quad
\hat{\cH}^{reg}\,|\phi\ra=0
\quad\Rightarrow\quad
\hat{\cH}^{reg}\,\hat{\cC}\,|\phi\ra=0
\,.
\ee
So $\hat{\cC}$ also acts on the space of physical states. This is the quantum counterpart of the classical fact that $\cC$ is constant along trajectories.

%In this formulation mixing both matter and geometry, it is not clear a priori how to deparametrize the quantum theory and extract semi-classical trajectories or a time evolution from the physical states. On the other hand, 

 We have therefore obtained a clean quantization which self-adjoint operators representing both the Hamiltonian constraint and the complexifier and thus generating unitary flows on the Hilbert space.
A second step should be to investigate how to deparametrize the quantum theory and extract semi-classical trajectories. In particular, defining a suitable Dirac's observable for the volume and compute its spectrum would complete the study of this model. We will address those crucial questions in a next paper.

%%%%%%%%%%%%%%%%
\section{Discussion}
\label{discuss}
%%%%%%%%%%%%%%%%

In this work, we have studied the Immirzi generator $\cC$ in cosmology at both classical and quantum levels, as well as the CVH Poisson algebra it forms with the volume and the Hamiltonian scalar constraint. The interest in considering the CVH algebra is that it exhibits a $\su(1,1)$ Lie algebra structure which encodes several properties of the classical theory. Of particular interest for us is that it encodes the fact that the BI parameter is not a physical parameter at the classical level, since the scalar constraint is simply globally rescaled under the Immirzi generator flow, i.e. $\{ \cC, \cH\} = - \cH$. Moreover, because of its $\su(1,1)$ Lie algebra structure, it  provides a suitable structure for a group quantization, free from factor-ordering ambiguities. Finally, the CVH algebra refers to the full theory, gravity plus matter. Thus, preserving this structure in the quantum theory provides a quantization scheme free from factor ambiguity and in which the BI parameter can be rescaled freely without affecting the dynamics and the physical predictions, i.e. here the Immirzi ambiguity disappears.

 It is nevertheless important to keep in mind that our framework distinguishes the Barbero-Immirzi parameter $\gamma$, resulting from the canonical transformations generated by the complexifier, from the  regularization scale $\lambda$ entering the Hamiltonian constraints. This is the main difference with the standard formulation of loop quantum cosmology, in which these two parameters are identified. Here, on the contrary, the regularization scale $\lambda$ is a physical parameter, while the Barbero-Immirzi parameter $\gamma$ is shifted by unitary transformations which do not affect $\lambda$.

 More precisely, we have presented a new LQC-inspired model which preserved the structure of the CVH algebra. In order to do so, we have generalized the regularization procedure of standard LQC. We have introduced a regularization of the complexifier, as well as a regularization of the inverse volume factor, in order to keep a closed CVH algebra. This regularization mostly account for updating the connection to a holonomy in the expression of the complexifier, which is consistent with the prescription used for the Hamiltonian constraint. As a consequence, the regularized volume factor also inherits a correction term depending on the connection $b$.
In this context, one can proceed to the group quantization of the model. It results in a quantum theory where the transformations generated by the regularized Thiemann complexifier are unitary,. This means that the BI parameter can be rescaled through unitary transformations, thereby solving  the Immirzi ambiguity in this very simplified quantum mechanical model.

Preserving the classical CVH algebra structure allows also to export some appealing properties of the classical theory in our new LQC model. Indeed, at the classical level, the deparametrized cosmological evolution, with respect to the scalar field, is given precisely by the complexifier. Our new regularization preserves this feature in the effective theory as well as in the quantum theory. The fact that the relational cosmological evolution can be generated by the generator of geometry rescaling suggests that there is an interplay between the scale transformations of the geometry, and the dynamics as seen from the relational clock, if not an identification. This observation begs the question concerning the role of the extrinsic curvature as a generator of the dynamics in symmetry reduced loop models and more generally in LQG, as already noticed for Gowdy models in a specific gauge as shown in \cite{Marugan1, Marugan2}. 
%
%In this line of though, it is interesting to note that the complexifier, given as the trace of the extrinsic curvature, is actually the Gibbons-Hawking boundary term added to the Einstein-Hilbert action in order for the action to be differentiable. It is well known fact that for a bounded region of spacetime, the scalar constraint is equal to the two dimensional integral of the extrinsic curvature, which gives rise to the quasi-local notion of energy introduced by Brown and York \cite{Brown, Regge}. This point underlines the potential link between the dynamics in the bulk as generated by the scalar constraint, and a boundary dynamic that would be generated by the complexifier. This idea is especially appealing in the context of the recently introduced boundary states as a coarse graining of the bulk spin network proposed in \cite{EH}. We plan to investigate this in a future work.

\smallskip

 Another important point of our framework is that the CVH algebra and the $\SU(1,1)$ structure it generates allows to integrate exactly the action of the Hamiltonian constraint and of the complexifier, at the classical and quantum levels. This must necessarily be compared to the existing exactly solvable version of loop quantum cosmology, as introduced in \cite{sLQC}. The main difference is the regularization of the complexifier and of the inverse volume factor. It is true that the Hamiltonian constraints only differ by a overall factor. This modification, which can be re-absorbed in the definition of the lapse and thus can not change the physics at the classical level, nevertheless affects slightly the quantization of the constraint and allows to get rid of all factor ordering ambiguities under the condition that the CVH algebra is preserved at the quantum level. We can summarize these differences in the table below.
 \begin{table}[H]
\centering
\begin{tabular}{ | c || c | c | c | c | c | c | c |  }
 		\hline
 		 & \quad CVH algebra \, $$  & \quad Input from the full theory \, $$ & \quad Immirzi rescaling \, $$ & \quad Factor ordering issues \,  $$ \\
 		\hline \hline
 		\, sLQC \, $$  &  broken & kinematical area spectrum & non unitary & present  \\
 		\,  New model\, $$ &  preserved  & none & unitary & none  \\ 
 		\hline
 	\end{tabular}
	\caption{ Comparative table on different properties of the sLQC model and the new LQC-inspired model introduced in the present work. \label{tableI2M+3M}}
 \end{table}

\medskip

Let us come back to the Immirzi ambiguity. The main result on this issue is that, thanks to the $\su(1,1)$ structure of the CVH algebra, one can build a LQC-inspired model where the BI parameter doesn't play any physical role in the quantum theory. In particular, it should not be understood as a new fundamental constant of the model. The Barbero canonical transformation can be mapped to a unitary transformation at the quantum level in this construction, solving thus the Immirzi ambiguity present in standard LQC. While our quantization refers to a very simplified (quantum mechanical) cosmological system, the result obtained within this framework  provides a first hint on the status of the Barbero-Immirzi parmeter. Whether this conclusion can be generalized to more complex systems, such as the loop quantization of spherical symmetric space-time or isolated horizons need to be investigated. We plan to address those questions in future work.

Finally, we have seen that the CVH algebra represents a powerful structure which allows a straightforward quantization of the homogenous and isotropic loop phase space. It would be particularly interesting to investigate this structure beyond homogeneity, at the level of the cosmological perturbations. Indeed, the treatment of the cosmological perturbations in LQC has been developed along two very different lines, the deformed algebra approach \cite{DefAlgebra} and the dressed metric approach \cite{DressMetric}. While the former is purely effective, the second provides a quantum theory of the perturbations over a quantum background, an effective description being derived from it. If the CVH algebra structure survive the truncation at second order, one could investigate the holonomy corrections required to preserve its classical structure, similarly to what we have done at in section II. It could potentially provide a starting point for a group quantization of the deformed algebra approach to the cosmological perturbations in LQC and allow to investigate at the quantum level the special features arriving from this covariant approach, such as the signature change phenomenon. See \cite{SG1, SG2} for details on this point. Yet, if the CVH algebra fails to close beyond homogeneity, it could signal interesting sign of conformal anomalies. We leave this idea for future investigations.

Moreover, the framework developed in this paper represents also a suitable formalism to apply the Wick rotation program proposed in \cite{ThiemannRC} to LQC, from the standard Ashtekar-Barbero to the self dual initial formulation. Since this idea motivated initially the work presented here, let us briefly mention the reasons for developing the self dual version of LQC. Recent investigations have shown that the standard strategy used in polymer quantization of symmetry reduced loop models can lead to inconsistencies, due to the introduction of holonomy corrections. While symmetry reduced model without local degrees of freedom exhibit a generic deformed notion of covariance \cite{BojDef, BojDefBH}, the holonomy corrected models with local degrees of freedom present anomalies and fail to be covariant. Those no go results were worked out in \cite{BojSS} for spherically symmetric model coupled to matter, and in \cite{BojGowdy} for polarized Gowdy model. They provide strong obstructions to the development of symmetry reduced loop models with local degrees of freedom, and in particular to black hole models and to the study of Hawking radiation in such models \cite{G1}. However, all those conclusions were derived within the Ashtekar-Barbero formulation. Even more recently, it was shown that those obstructions disappear when working within the initial self dual formulation. In such self dual symmetry reduced loop models, the holonomy corrected hypersurface deformation algebra reproduces without deformation its classical counterpart, both for model without or with local degrees of freedom \cite{SDHypSS, SDHypCP}. It indicates therefore that self dual variables are better suited to preserve covariance in presence of holonomy corrections in such models. In particular, it has been shown in \cite{SDHypCP} that the inhomogenous self dual LQC model does not suffer from the signature change deformation present within the Ashtekar-Barbero formulation. Because of this recent observation, it is crucial to develop the self dual version of LQC, and investigate its inhomogeneous extensions, since it provides an inequivalent approach to the cosmological perturbations in LQC, with potentially different predictions. 
%Additionally, it was shown recently that the bouncing dynamic of the homogenous and isotropic background in LQC is equivalently described by the cosmological  sector of a scalar tensor theory \cite{TSLQC1, TSLQC2}, given by the Chamseddine-Muhkanov mimetic theory \cite{CM}. While this equivalence holds at the homogenous and isotropic level, it is not clear if it can be extended beyond homogeneity. (See \cite{CPMimetic} for investigation in inhomogeneous mimetic non singular cosmology). The reason is that while the Chamseddine-Muhkanov theory,  as any covariant theory of gravity, exhibits the same hypersurface deformation algebra as General Relativity, the inhomogeneous LQC model obtained within the deformed algebra approach posses a deformed notion of covariance which cannot be absorbed by redefinition of the lapse \cite{BojHyp}. On the contrary, the approach based on the self dual variables reproduce exactly the hypersurface deformation algebra of classical GR. Therefore, the self dual formulation could represent a better candidate to build a mapping with a covariant scalar-tensor theory as proposed in \cite{TSLQC1, TSLQC2} which holds beyond homogeneity.  
For the different reasons advocated here, developing the self dual version of LQC is mandatory and the formalism developed in this paper represent the natural framework for this purpose. One could then compare the model obtained with the existing proposals in the literature such as  \cite{EdSD1, EdSD2, AnaLQC}.

%%%%%%%%%%%%%%%%%%%%%%%%%%%%%%%%%
%%%%%%%%%%%%%%%%%%%%%%%%%%%%%%%%%
\appendix

%%%%%%%%%%%%%%%%
\section{The CVH algebra of classical General Relativity}
\label{CVHgen}
%%%%%%%%%%%%%%%%

In this section, we compute the CVH algebra in full General Relativity. In term of the Ashtekar-Barbero variables, General Relativity can be formulated as an $\SU(2)$ gauge field theory. This phase space is coordinatized by the two conjugated variables, the Ashtekar-Barbero connection $A^i_a dx^a$ and its momentum, the densitized electric field $E^a_i \partial_a$, which admit the following Poisson bracket
\be
\{ A^i_a(x), E^b_j(y)\} = \kappa \gamma \delta^b_a \delta^i_j \delta^{(3)}(x-y)
\ee

Those variables are build from the tetrad field $e^i_a dx^a$, the extrinsic curvature $K^i_a dx^a$ and the rotational spin connection $\Gamma^i_a dx^a$ as
\be
E^a_i = \frac{1}{2} \epsilon^{abc} \epsilon_{ijk} e^j_b e^k_c = \text{det}(e) e^a_i \qquad \qquad A^i_a = \Gamma^i_a + \gamma K^i_a
\ee

The canonical variables $(A^i_a,E^b_j )$ are subject to seven first class constraints (after canonical analysis and having solve the second class constraints showing up in the process) given by the Gauss constraint $\cG_i$, the (spatial) diffeomorphism constraint $\cH_a$ and the scalar constraint $\cH$ enforced respectively by the lagrange multipliers $A^i_0$, $N^a$ and $N$ 
\begin{align}
& \cG_i =  \partial_a E^a_i + \epsilon_{ij}{}^k A^j_a E^a_k \qquad  \cH_a = E^b_i F^i_{ab} \qquad  \cH = - \frac{1}{2\gamma^2} \frac{E^a_i E^b_j}{\sqrt{\text{det}(E)}} \; ( \; \epsilon^{ij}{}_k  F^k_{ab}  - 2 (1 + \gamma^2) K^{[i}_a K^{j]}_b \; )
\end{align}
It is interesting to consider the euclidean and Lorentzian contribution to the scalar constraint separatly. We introduce thus the notation
\be
\cH = \cH^{E} + \cH^K \qquad \cH^E = - \frac{1}{2\gamma^2} \frac{E^a_i E^b_j}{\sqrt{\text{det}(E)}} \; \epsilon^{ij}{}_k  F^k_{ab}   \qquad \cH^K = \frac{(1 + \gamma^2)}{\gamma^2} \frac{E^a_i E^b_j}{\sqrt{\text{det}(E)}} \; K^{[i}_a K^{j]}_b 
\ee
The spin connection turns out to be related to the electric field $E^a_i$ by the following expression
\begin{align}
\Gamma^i_a & = \frac{1}{2} \epsilon^{ijk} E^b_k\; \big{[} \; \partial_b E^j_a - \partial_a E^k_b + E^c_j E^l_a \partial_b E^l_c \; \big{]}   + \frac{1}{4} \epsilon^{ijk} E^b_k \;\big{[} \; 2 E^j_a  \frac{\partial_b \text{det}(E)}{\text{det}(E)} - E^j_b \frac{\partial_a \text{det}(E)}{\text{det}(E)}  \; \big{]}
\end{align}
Let us now introduced the following functions on the phase space
\be
 \cC = \frac{1}{\gamma \kappa} \int_{\Sigma} d^3x \;  E^a_i K^i_a \qquad  V = \int_{\Sigma} d^3 x \sqrt{ \frac{1}{3!} \epsilon^{ijk} \epsilon_{abc} E^a_i E^b_j E^c_k} \qquad \cH = \cH^E + \cH^K
\ee
A useful bracket for our purpose is given by
\be
\{ \cC, \Gamma^i_a\} = 0 
\ee
%where we have introduce the rescaled extrinsic curvature $\tilde{K}^i_a = \gamma K^i_a$. Expressing the complexifier using the (rescaled) extrinsic curvature instead of the Ashtekar-Barbero connection will considerably simplify the computation. Recall that $\tilde{K}^i_a$ and $A^i_a$ are related through a canonical transformation and therefore share the same Poisson bracket with the electric field $E^a_i$
%\be
%\{ \tilde{K}^i_a(x), E^b_j(y)\} = \gamma \kappa \delta^b_a \delta^i_j \delta^{(3)}(x-y) \qquad \text{while} \qquad  \{ \tilde{K}^i_a(x), \tilde{K}^b_j(y)\} = 0
%\ee
Let us now compute the (CVH) algebra generated by the three phase space functions introduced above. The brackets read
\begin{align*}
& \{ \cC , V \}  =   V\qquad \{ \cC, \cH^E\} =  -\cH^E   \qquad  \{ \cC, \cH^K \} = - \cH^K   \qquad \{ V, \cH^E\} = \cC \qquad \{ V, \cH^K \} = 0 
\end{align*}
This the CVH algebra of classical General Relativity.

%Next, we consider the euclidean part of the scalar constraint $H^E$. Ignoring the determinant which can always be removed by a suitable choice a the lapse $N$, it reads
%\be
%\cH^{E} = \frac{1}{\gamma^2} \epsilon^{abc} \epsilon_{ijk} E^a_i E^b_j B^c_k \qquad \text{with} \qquad B^c_k = \frac{1}{2} \epsilon^{cef} F_{ef \; k}
%\ee
%The Poisson bracket between the euclidean scalar constraint and the volume leads to
%\be
%\label{HV}
%\{ \cH^{E} , V \} =  \epsilon^{abc} \epsilon_{ijk} E^a_i E^b_j \{ B^c_k , V\}
%\ee 
%Using that the bracket between the connection $A^i_a$ and the volume gives us exactly the first term of the l.h.s  of \ref{HV} as
%\be
%\{ A^k_c , V \} =  3\kappa \gamma \;  \epsilon^{cde} \epsilon_{kmn} E^m_d E^n_e 
%\ee

%we obtain the following form for the bracket \ref{HV}
%\be
%\{ \cH^E, V \} \propto \frac{1}{ \gamma^3 } \;  \{ A^k_c , V \}  \{ B^c_k , V\}
%\ee

%One can then show that 
%\be
%\cC = \frac{1}{\gamma} \int dx^3 A^i_a E^a_i = \int dx^3 \; \{ \cH^E, V \} 
%\ee

%%%%%%%%%%%%%%%%
\section{The  $\SU(1,1)$ action on the $(v,b)$ phase space}
%%%%%%%%%%%%%%%%

%%%%%%%%%
\subsection{The spinor phase space and the $\SU(1,1)$ action }
%%%%%%%%%

Following  the group theoretic reformulation of loop quantum cosmology in the deparametrized formulation in terms of a $\SU(1,1)$ symmetry introduced in \cite{LivGQCosmo}, we find convenient to use a spinorial presentation of the $\su(1,1)$ Lie algebra to exponentiate the action of the generators and later to define proper coherent states.
We introduce a pair of canonical complex variables, $z^{0}$ and $z^{1}$, provided with the canonical Poisson bracket:
\be
\label{CRR}
\{z^0, \bar{z}^0\} = \{ z^1, \bar{z}^1 \} = - i \,.
\ee
On this phase space, one can identify a representation of the $\su(1,1)$ Lie algebra with generators:
\begin{align}
\label{zz}
j_{z} = \frac{1}{2} ( |z^{0}|^{2} + |z^{1}|^{2})
% = v
\,,\qquad
k_{+} = \bar{z}^{0} \bar{z}^{1}
%= v e^{2i \lambda b}
\,,\qquad
k_{-} = z^{0} z^{1}
%= v e^{-2i \lambda b} 
\,,
\end{align}
which satisfy the $\su(1,1)$ commutation relations
\be
\{ j_z,k_{\pm} \} = \mp i k_{\pm}\,, \qquad \{ k_{+}, k_{-}\} = 2 i j_z
\,.
\ee
We can also define the real boost generators:
\be
k_{x}=\f12(k_{+}+k_{-})
\,,\qquad
k_{y}=\f1{2i}(k_{+}-k_{-})
\ee
We can also introduced another observable $L$ which commutes with the previous generators
\be
\ell = \frac{1}{2} ( |z^{0}|^{2} - |z^{1}|^{2}) \qquad \{\ell, j_z\} = \{ \ell, k_{\pm} \} = 0\,.
\ee
This Casimir observable can actually be understood as the  square root of the quadratic Casimir of the $\su(1,1)$ algebra:
\be
\mathfrak{C} = k_{+}k_{-}-j^2_z = -\ell^2 \le 0\,.
\ee
As we see, this spinorial representation only allows for time-like representation with negative or vanishing Casimir.
It is nevertheless to change the definition of the $\su(1,1)$ generators to reach all possible unitary representations of $\SU(1,1)$.

It is convenient to re-package the generators as a 2$\times$2 Hermitian matrix:
\be
\mathfrak{m}
\equiv
\mat{c}{z^{0} \\ \bz^{1}}
\mat{c}{z^{0} \\ \bz^{1}}^{\dagger}
=
\mat{cc}{|z^{0}|^{2} & z^{0} z^{1} \\ \bz^{0} \bz^{1} & |z^{1}|^{2} }
\,,\qquad
M\equiv\mathfrak{m}-\f12\big{(}\tr\,\mathfrak{m}\sigma_{z}\big{)}\,\sigma_{z}=
\mat{cc}{j_{z}&k_{-}\\ k_{+}&j_{z}}\,.
\ee
We can compute the action of the $\su(1,1)$ generators on the spinor $z$ and thus also on the matrix $M$. Introducing $\veta\cdot\vj=\eta_{z}j_{z}-\eta_{x}k_{x}-\eta_{y}k_{y}$ using the Lorentzian signature, we have:
\be
\Big{\{}\veta\cdot\vj,\,\mat{c}{z^{0}\\ \bz^{1}}\,\Big{\}}
=
\f i2 \veta\cdot\vtau\,\mat{c}{z^{0}\\ \bz^{1}}
\,\qquad
{\{}\veta\cdot\vj,\,\mathfrak{m}\,{\}}
=
\f i2\bigg{(}\veta\cdot\vtau\,\mathfrak{m}-\mathfrak{m}\,\veta\cdot\vtau^{\dagger}\bigg{)}\,,
\ee
where the $\tau_{a}$ are the Lorentzian Pauli matrices:
\be
\tau_{z}=\mat{cc}{1 & 0 \\ 0 & -1}=\sigma_{z},
\quad
\tau_{x}=\mat{cc}{0 & 1 \\ -1 & 0}=+i\sigma_{x},
\quad
\tau_{y}=\mat{cc}{0 & -i \\ -i & 0}=-i\sigma_{y}\,.
\ee
Taking into account that the rotation generator $\tau_{z}^{\dagger}=\tau_{z}$ is Hermitian while the boost generators, $\tau_{x}^{\dagger}=-\tau^{x}$, $\tau_{y}^{\dagger}=-\tau^{y}$, are anti-Hermitian, we deduce the transformation law for the generator matrix $M$:
\be
{\{}\veta\cdot\vj,\,\tr\,\mathfrak{m}\sigma_{z}\,{\}}
=
0\,,\qquad
{\{}\veta\cdot\vj,\,M\,{\}}
=
\f i2\bigg{(}\veta\cdot\vtau\,M-M\,\veta\cdot\vtau^{\dagger}\bigg{)}\,.
\ee
This can be directly exponentiated to get the flow generated by the $\vj$'s on the phase space:
\be
e^{\{\veta\cdot\vj,\cdot\}}\,
\mat{c}{z^{0}\\ \bz^{1}}
=G_{\veta}\,\mat{c}{z^{0}\\ \bz^{1}}
\,,\quad
e^{\{\veta\cdot\vj,\cdot\}}\,
M
=G_{\veta}\,MG_{\veta}^{\dagger}
\,,\qquad\textrm{with}\quad
G_{\veta}=e^{\f i2 \veta\cdot\vtau} \in\SU(1,1)\,.
\ee

%%%%%%%%%
\subsection{The $(v,b)$ phase space from the spinor variables}
%%%%%%%%%

Since the observable $\ell$, measuring the difference in norm between the two complex variables, commutes with the $\su(1,1)$ generators, we can fix the value of this Casimir without affecting the transformation law under the $\SU(1,1)$-action.
Let us start with the vanishing Casimir case, $\ell=0$. In this case, the two complex variables have equal norms, $|z^{0}|=|z^{1}|$, and we define the variable $v\ge 0$ as that norm and introduce the relative phase $b$ between them:
\be
\f v{2\lambda}=\f12(|z^{0}|^{2}+|z^{1}|^{2})
\,\quad
e^{-2i\lambda b}=\f{z^{0}}{\bz^{1}}
\,\qquad
z^{0}=\sqrt{\f v{2\lambda}}e^{-i\lambda b+i\varphi}
\,\quad
z^{1}=\sqrt{\f v{2\lambda}}e^{-i\lambda b-i\varphi}\,,
\ee 
where $\lambda$ is a fixed constant and $\varphi$ is an arbitrary  phase variable. We can check that $\ell$ commutes with both $v$ and $b$, so that they define proper coordinates on the constrained surface $\ell=0$. And we further check that they form a canonical pair of variables:
\be
\{\ell, v\}=\{\ell, e^{-2i\lambda b}\}=0
\,,\quad
\{\f v{2\lambda},e^{-2i\lambda b}\}=i\,e^{-2i\lambda b}
\qquad\Rightarrow\quad
\{b,v\}=1
\,.
\ee
In that case, we compute the $\su(1,1)$ generators in terms of this parametrization of the complex variables. They do not depend  on the phase $\varphi$ and are expressed simply in terms of the $(v,b)$ variables:
\be
j_{z}=\f v{2\lambda}\,,\qquad k_{\pm}=\f v{2\lambda}\,e^{\pm2i\lambda b}\,.
\ee

\smallskip

We can generalize this analysis to the case of a non-vanishing Casimir $\ell\ne 0$. Let us consider $\ell=(2\lambda)^{-1}v_{m} >0$. The case $\ell<0$ can be treated similarly. We take the exact same definition for the volume $v$ and the phase $b$, which still form a canonical pair of variables which commute with $\ell$.  However, the expression of the complex variables changes slightly to accommodate the difference in norms:
\be
z^{0}=\sqrt{\f {v+v_{m}}{2\lambda}}\,e^{-i\lambda b+i\varphi}
\,\quad
z^{1}=\sqrt{\f {v-v_{m}}{2\lambda}}\,e^{-i\lambda b-i\varphi}\,,
\ee
which leads to the following expression for the $\su(1,1)$ generators:
\be
j_{z}=\f v{2\lambda}\,,\qquad k_{\pm}=\f {v^{2}-v_{m}^{2}}{2\lambda}\,e^{\pm2i\lambda b}
\,\qquad
\mathfrak{C}=-\ell^{2}=-v_{m}^{2}<0
\,.
\ee
This provides the regularization of the $(v,b)$ phase space to account for the existence of a minimal volume $v_{m}$ at the kinematical level and its embedding in the $\su(1,1)$ phase space.

%%%%%%%%%
\subsection{Space-like Representations of $\SU(1,1)$}
\label{newsu11}
%%%%%%%%%

Let us look into the deformation of the $\su(1,1)$ generators due to the matter Hamiltonian.
Start with $\su(1,1)$ algebra generators $\vj$ and define new generators:
\be
K_{y}=k_{y}
\,,\qquad
K_{x}=k_{x}+\f\beta{2(j_{z}+k_{x})}
\,,\qquad
J_{z}=j_{z}-\f\beta{2(j_{z}+k_{x})}
\ee
\be
\{J_{z},K_{x}\}=K_{y}
\,,\qquad
\{J_{z},K_{y}\}=-K_{x}
\,,\qquad
\{K_{x},K_{y}\}=-J_{z}
\,,
\ee
\be
\mathfrak{C}=J_{z}^2-K_{x}^2-K_{y}^2=\big{[}j_{z}^2-k_{x}^2-k_{y}^2\big{]}-\beta\,.
%\widetilde{\mathfrak{C}}=
%\mathfrak{C}+\beta\,.
\ee
In our framework for cosmological models, the initial Casimir  $j_{z}^2-k_{x}^2-k_{y}^2$ vanishes, so a positive $\beta>0$ produces a negative $\su(1,1)$-Casimir, leading to a space-like representations.
We can introduce the following presentation of $\su(1,1)$ in terms of spinors, which allows for space-like representations and which is different from the one used for time-like representations:
\be
J_{z}=\f12(|z^0|^2-|z^1|^2)
\,,\qquad
K_{+}=\f12((\bz^0)^2-(z^1)^2)
\,,\qquad
K_{-}=\f12((z^0)^2-(\bz^1)^2)
\ee
\be
\{J_{z},K_{\pm}\}=\mp i K_{\pm}
\,,\qquad
\{K_{+},K_{-}\}=2iJ_{z}
\ee
\be
\mathfrak{C}=J_{z}^2 - K_{+}K_{-}
=-\lambda^2
%= - (s^2 + \frac{1}{4})
< 0
\,,\qquad
\lambda =\f i2(\bz^0\bz^1-z^0z^1)\in\R\,.
\ee
%where we have introduce the continuous parameter $\lambda = i s - 1/2$ for simplicity. The continuous representation of $\SU(1,1)$ are labelled by two parameter, the continuous spin $s$, or equivalently by the complex continuous spin $\lambda$, and by a discrete magnetic number $\mu$. There is two sub-class of in the continuous series denoted $\cC^{\epsilon}_\lambda$, in which the parameter $\mu$ is slightly different. For $\epsilon = 0$, we obtain
%\be
%\lambda = - \frac{1}{2} + is \qquad 0 < \infty < + \infty \qquad \mu = 0, \pm 1, \pm 2, ...
%\ee
%while for $\epsilon = 1$, we get
%\be
%\lambda = - \frac{1}{2} + is \qquad 0 < \infty < + \infty \qquad \mu = \pm \frac{1}{2}, \pm \frac{3}{2}, ...
%\ee
%Since the magnetic number $\mu$ corresponds to the volume eigenvalues in our framework, it is likely the second sub-class which should play a role, since the zero eigenvalue doesn't belong to the volume spectrum in that case. This is likely the correct framework to define $\SU(1,1)$ coherent states.

Canonically quantizing the two complex variables, $\{z^0,\bz^0\}=\{z^1,\bz^1\}=-i$, as a pair of harmonic oscillators allows to recover all the space-like $\SU(1,1)$-representations $\cC^\eps_{s}$, which have a negative Casimir. The operator $\hJ_{z}$ becomes half the difference of energy between the two oscillators, it has a discrete spectrum and its eigenvalues runs on all possible positive and negative half-integers $m\in\Z/2$. Using this spinorial formulation would allow to define $\SU(1,1)$-coherent states for those space-like representation, similarly to what has been developed for the time-like representation (from the discrete series) in \cite{LivGQCosmo}.

%%%%%
\section{Introducing a minimal volume $v_m$ in the classical phase space}
%\subsection{On the additional volume Regularization and the new scale $v_m$}
\label{fullreg1}
\label{vmin}
%%%%%

{As already pointed out in \cite{LivGQCosmo}, the vanishing $\sl_{2}$ Casimir of the gravitational sector can be naturally regularized into a non-zero Casimir  by accounting for a minimal volume at the classical level. Indeed, a  generalization of the definition of the $\su(1,1)$ generators in terms of the $(v,b)$ variables is:
\be
j_{z}=(2\lambda)^{-1}v
\,,\qquad
k_{\pm}=\sqrt{v^{2}-v_{m}^{2}}\,e^{\pm  2\lambda i}
\,,
\ee
where we have modified the boost generators by introducing the new parameter $v_{m}$. This sets a minimal value for the volume, which can be identified to the {\it volume gap} of LQC. The volume $v$ can vanish no more and it is necessarily larger than $v_{m}$.
The $\su(1,1)$ Lie algebra is not modified, but its Casimir does not vanish anymore:
\be
\label{CazLQCVol}
\mathfrak{C}_{\sl_{2}}=j_{z}^{2} - k_{+}k_{-}= v_{m}^{2} > 0\,.
\ee
At the quantum level, this will select a time-like representation of $\SU(1,1)$, chosen by the minimal value of the volume and whose lowest weight vector will be the minimal volume state.

This extension leads to a generalized classical Hamiltonian constraint:
\beq
\label{RegVol}
\cH^{m}
&=&
\frac{\pi^2_{\phi}}{24 \pi G v}
- \frac{1}{2} \sqrt{v^{2}-v_{m}^{2}} \frac{\sin^2{(\lambda b)}}{\lambda^2}
-\f v{4\lambda^{2}}\left(1-\sqrt{1-\f{v_{m}^{2}}{v^{2}}}\right)\nn\\
&=&
\cH-\f1{8\lambda^{2}}\f{v_{m}^{2}}v\,\cos2\lambda b +\dots
\eeq
This generalization naturally takes into account the volume gap of loop quantum cosmology and maintains the integrability of the motion based on the $\SU(1,1)$ group structure. The complexifier also gets slightly modified:
\be
\cC^{m}=k_{y}=\sqrt{v^{2}-v_{m}^{2}}\,\f{\cos 2\lambda b}{2\lambda}\,,
\ee
but its flow on the $\su(1,1)$ generators $\vj$ remains the same. 

Finally, we can put together the $\lambda$-regularization with the minimal volume regularization:
\beq
\label{Hfull}
\cH^{full}
&\equiv&
\frac{\pi^2_{\phi}}{24 \pi G \lambda(k_{x}+j_{z})}
+\f1{2\lambda}(k_{x}-j_{z})
\quad\textrm{with}\quad
j_{z}
=\f v{2\lambda}\,,
\,\,
k_{x}=\sqrt{v^{2}-v_{m}^{2}}\f{\cos2\lambda b}{2\lambda}
\nn\\
&=&
\frac{\pi^2_{\phi}}{12 \pi G \big{[}v+\cos2\lambda b\sqrt{v^{2}-v_{m}^{2}}\big{]}}
-\f1{2\lambda^{2}}
\bigg{[}
\sqrt{v^{2}-v_{m}^{2}}\,\sin^{2}\lambda b+\f 12(v-\sqrt{v^{2}-v_{m}^{2}})
\bigg{]}
%-\f1{4\lambda^{2}}
%\bigg{[}
%v-\sqrt{v^{2}-v_{m}^{2}}\,\cos2\lambda b
%\bigg{]}
\eeq
This is the most general ansatz for an effective loop cosmology Hamiltonian leading to a closed CVH algebra in our framework and accounts of both curvature regularization by $\lambda$ and the existence of a volume gap $v_{m}$. The action of the regularized complexifier $\cC$  acts non-trivially on the variables $v$ and $b$, but  simply rescales the whole Hamiltonian constraint without changing its inner structure.

\smallskip

This way of introducing a minimal volume is rather different from what is commonly done in the LQC literature.
Both approaches share the same regularization of the curvature. However, the existence of a minimal volume (or  of an area gap) is introduced in a very different way in the LQC literature and in the present approach. In the LQC literature \cite{LQCReport}, the existence of an area gap is introduced as an ad hoc input, imported from the full LQG theory. The value of the area gap $\Delta$ is directly imported from the kinematical area spectrum, which is $\gamma$-dependent. This value is then used at the dynamical level in LQC. However, this spectrum has been computed in the full theory only at the kinematical level, and using  it at the dynamical level in a symmetry reduced model as LQC can be questionable. On the contrary, in our approach, the minimal volume is introduced through an additional step in the regularization leading to the effective phase space to quantize. In the present framework, we have imported no ingredient from the full LQG theory and focused on the regularized phase space for cosmology on its own, exploring its ambiguities and consistent extensions.  As a result, the Barbero-Immirzi parameter $\gamma$, the regularization scale $\lambda$ and the minimal volume $v_{m}$ are a priori independent and their effect and physical relevance can be studied separately.

\medskip

What is especially interesting in this approach is that the $\sl_{2}$ Casimir now gets opposite contributions from both regularizations by $\lambda$ and $v_{m}$:
\be
\label{CazLQCCMatVol}
\mathfrak{C}_{\sl_{2}}
= - 2V\cH^{full} - \cC^{2}
=J_{z}^{2} - K_{x}^{2}- K_{y}^{2}
=\f{\pi_{\phi}^{2}}{12\pi G}+\big{[}j_{z}^{2} - k_{x}^{2} - k_{y}^{2} \big{]}
=v_{m}^{2} - \f{\pi_{\phi}^{2}}{12\pi G}\,.
\ee
This Casimir equation shows a competition between quantum gravity effects and the matter sector which determine the Hilbert space to use for the quantum theory. It is a remarkable fact that both  gravity and matter sectors are involved at the same fundamental level in the construction of the quantum theory. 
Depending on the relative values of $v_{m}$ and $\pi_{\phi}$, the Casimir will be either positive or negative, thus leading respectively to a choice of time-like or space-like representation. Mathematically, both are possible. Physically, the minimal volume will be at the Planck scale and thus the matter energy will dominate the expression, thus giving a negative Casimir and leading to a quantization using a representation of the continuous series.

The formula for the Casimir (\ref{FullCaz}) suggests that there may be a more general underlying $\SL(2,\mathbb{C})$ structure to elucidate. Indeed,  the Casimir for $\SL(2,\C)$-representations is $\mathfrak{C}_{\sl(2,\C)} = \frac{1}{2} (n^2 - \rho^2 - 4)$ where $n \in \mathbb{Z}^{+}$ and $\rho \in \mathbb{R}$. Matching this against our formula, it is tempting to identify the volume $v_m$ as the discrete contribution to the ${\sl(2,\C)}$-Casimir $n^2$, while the continuous negative term $- ( \rho^2 + 4)$ would correspond to the matter contribution. Such a $\SL(2,\mathbb{C})$ complexified structure would be likely also very relevant to a Wick-rotation to imaginary values of the BI parameter.

%%%%%%%%%%%%%%%%%%%%%%%%%%%%
%%%%%%%%%%%%%%%%%%%%%%%%%%%%
\bibliographystyle{bib-style}

\bibliography{LQCI}

\end{document}